\documentclass[%
 preprint, 
 amsmath,amssymb,
 aps, physrev,
]{revtex4-2}

\usepackage{graphicx}
\usepackage{dcolumn}
\usepackage{bm}
\usepackage{physics}
\usepackage{xcolor}

\begin{document}

\preprint{APS/123-QED}

\title{\textbf{Soliton interaction and bound state formation in coupled Kerr resonators} }%

\author{D. A. Dolinina}
\affiliation{Weierstrass Institute, Mohrenstr. 39, 10117 Berlin, Germany}
\author{D. V. Turaev}
\affiliation{Imperial College, London SW7 2AZ, United Kingdom}
\author{A. G. Vladimirov}
\affiliation{National Research University Higher School of Economics, 25/12 Bolshaya Pecherskaya Ulitsa, 603155 Nizhny Novgorod, Russia}

\date{\today}

\begin{abstract}
Soliton dynamics in coupled Kerr microcavities is an important aspect of frequency comb technologies, with applications in optical communication and precision metrology. We investigate a minimal system consisting of two nearly identical coupled Kerr microresonators, each operating in the soliton regime and driven by a separate coherent beam, and analyze the mechanisms that govern their soliton interactions. In the weak-coupling regime, the system supports multiple soliton clusters characterized by distinct soliton separations and stability. Numerical simulations indicate that asymmetric perturbations can alter soliton separations or destroy these states, while the imposed pump phase difference plays a key role in cluster selection. Together, these findings highlight previously unexplored regimes of dissipative soliton organization and suggest new strategies for controlling soliton ensembles in integrated photonic platforms.
\end{abstract}                              
\maketitle

\section{Introduction}

Kerr microcavities have emerged as a powerful platform for nonlinear optics, enabling the generation of optical frequency combs through soliton formation \cite{herr2014temporal,kippenberg2018dissipative}.
These microresonators, which rely on the third-order Kerr nonlinearity, support dissipative Kerr solitons  that propagate in a stable manner within the cavity \cite{herr2016dissipative}. Most studies of Kerr microcavity solitons are limited to investigating a single microcavity pumped by a continuous-wave (CW) injection, leaving the dynamics of coupled microcavities largely unexplored. On the other hand, multi-channel optical schemes using coupled Kerr microresonators offer a promising approach for integrating multiple optical components on a single chip, enabling advanced signal processing, tunable filtering, and all-optical switching in compact photonic circuits. In particular, a scheme involving two mutually coupled optical cavities was proposed to enhance energy conversion efficiency \cite{xue2019super}. Additionally, a new type of microcavity soliton has been observed in two coupled ring resonators with normal dispersion \cite{yuan2023soliton}.

Weak interactions between temporal cavity solitons are fundamental to the emergence of structured patterns and complex collective dynamics in a wide range of nonlinear dissipative systems. These interactions, typically mediated by overlapping soliton tails or long-range nonlocal perturbations, play a critical role in the formation of bound states (also known as `soliton molecules'), the self-organization of soliton ensembles, and the onset of symmetry-breaking phenomena. Two of the most widely studied theoretical frameworks for modeling such phenomena are various modifications of the complex Ginzburg–Landau equation (CGLE) \cite{Aranson2002} and the Lugiato–Lefever equation (LLE) \cite{Lugiato1987}. In optics the CGLE describes dissipative soliton interactions in mode-locked fiber lasers and other active media with gain and loss \cite{Akhmediev1997,akhmediev2005dissipative,soto2007soliton,Grelu2012,vladimirov1999numerical}, while the LLE governs temporal soliton interactions in coherently driven passive Kerr resonators \cite{parra2014dynamics,parra2017interaction,vladimirov2021dissipative}. In addition to the CGLE and LLE, short optical pulse interactions have recently been investigated, both numerically and analytically, through delay differential equation models of mode-locking \cite{nizette2006pulse,javaloyes2016dynamics,vladimirov2022short,vladimirov2023temporal}.

In these models, isolated temporal dissipative solitons emerge from a delicate balance between nonlinearity, dispersion (or diffraction), and energy exchange with the environment. When multiple solitons coexist, their weak interactions can lead to a variety of dynamical behaviors, including attraction, repulsion, bound state formation, oscillatory motion, and synchronization \cite{Malomed1991,Akhmediev1997,vladimirov2001stable,Descalzi2011,turaev2012long}. The nature of these interactions depends sensitively on system parameters and cavity boundary conditions, and can give rise to stable bound states with discrete equilibrium separations, often referred to as `soliton crystals' or `soliton molecules.' These structures have been observed in both fiber lasers \cite{Tang2005,Runge2015} and microresonators \cite{Herr2014,kippenberg2018dissipative}, and are of growing interest for applications in optical buffering, frequency comb stabilization, and ultrafast signal processing.

Despite extensive theoretical and experimental investigations, the regime of weak soliton interaction, where the overlap between individual solitons is small, remains a particularly subtle and analytically challenging limit. In this regime, the interaction forces become exponentially small, and the resulting dynamics are highly sensitive to small perturbations, such as periodic modulations or higher-order dispersion terms  \cite{turaev2012long,parra2017interaction, vladimirov2018effect,vladimirov2021dissipative}. Understanding the stability and controllability of weakly interacting soliton states is thus essential for the design of robust photonic systems based on dissipative soliton lattices or time-multiplexed cavity solitons.

In this paper, we study the interaction of solitons in two externally driven, coupled Kerr microcavities, modeled by a pair of coupled Lugiato-Lefever equations (LLEs) \cite{lugiato1987spatial,chembo2013spatiotemporal} and pumped by two coherent injection beams. This system can be derived from the coupled-cavity neutral delay differential equation model \cite{vladimirov2023neutral} and it incorporates key physical effects such as dispersion, Kerr nonlinearity, internal and external losses, external driving, and inter-cavity coupling. In the weak-coupling limit, we develop an asymptotic approach that reveals the existence and stability properties of distinct soliton clusters, characterized by a time-independent separation between the soliton coordinates. 

By performing numerical analysis of the soliton dynamics governed by these equations, we identify the conditions under which bound states and phase-locking phenomena arise \cite{tikan2021emergent}. We demonstrate that, consistent with our asymptotic analysis, varying the phase difference of the injected fields enables the existence of multiple stable soliton configurations. Furthermore, we investigate the impact of symmetry breaking induced by cavity length mismatch and show that tuning this mismatch provides a mechanism to control the soliton separation within a cluster. These results provide new insights into the controllability of soliton dynamics in photonic platforms and open avenues for the design of complex optical systems.

\section{Model equations}
We consider a system of two coupled ring resonators  pumped by two coherent light beams. In the mean-field approximation, this system is described by the following set of coupled LLEs:

\begin{gather}
    \partial_t A_1= -\dfrac{V}{2} \partial_\xi A_1 + i b \partial_{\xi\xi} A_1 + i |A_1|^2 A_1  -\left(1 + i\theta\right) A_1 + i\kappa A_2 + i\eta_1 e^{i\delta/2}, \label{eq:LLE_coupled1}\\
   \partial_t A_2= \dfrac{V}{2} \partial_\xi A_2 + i b {\partial_{\xi\xi} A_2} + i |A_2|^2 A_2  -\left(1 + i\theta\right) A_2 + i\kappa A_1 + i\eta_2 e^{-i\delta/2},
    \label{eq:LLE_coupled2}
\end{gather}
where $A_{1,2}(t,\xi)$ are the field envelopes, $t$ is the slow time counting the cavity roundtrips, $\xi$ is the retarded (fast) time,  $\kappa$ is the coupling coefficient, $\eta_{1,2}$ are the injection rates, $\delta$ is the phase difference between the pump fields in the two resonators, $b$ is the second-order dispersion coefficient, $\theta_{1,2}$ are the cavity detunings, and $V$ is a parameter proportional to the small length difference between the two resonators.
Using the approach of Ref.~\cite{vladimirov2023neutral}, the system (\ref{eq:LLE_coupled1})-(\ref{eq:LLE_coupled2}) can be obtained in the mean-field limit from a set of coupled neutral delay differential equations for two coupled cavities. This derivation implies that the field amplitudes must satisfy the periodic boundary conditions $A_{1,2}(t,-L/2)=A_{1,2}(t,L/2)$, where $L$ is the cavity round-trip time. In Eqs.~(\ref{eq:LLE_coupled1})-(\ref{eq:LLE_coupled2}), it is assumed that the two pumping beams originate from a single beam that is split into two parts, ensuring coherence and identical optical frequency. Note that the common phase $\phi$ of the two injections can be eliminated from Eqs.~(\ref{eq:LLE_coupled1})-(\ref{eq:LLE_coupled2}) through the variable transformation $A_{1,2} \to A_{1,2}\exp(i\phi)$. In particular, this transformation allows the imaginary unit in front of the injection terms to be removed from the equations.

\section{Asymptotic analysis of the soliton interaction}
\begin{figure}[thb]
\centering
\includegraphics[width=0.45\linewidth]{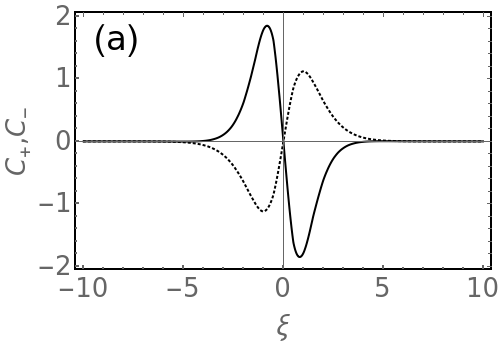}
\includegraphics[width=0.45\linewidth]{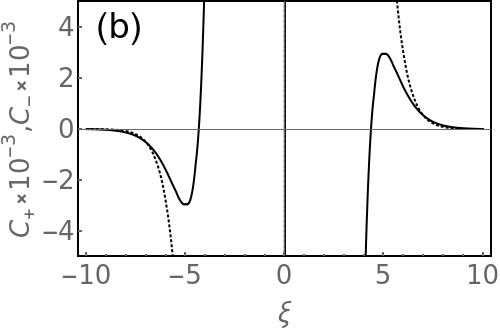}
\caption{a) Quantities $C_+$ (dotted line) and $C_-$ (solid line) as functions of the variable $\xi$.  
b) A magnified view of the neighborhood of zero from panel (a).  Parameter values: $\theta = 3.5$, $\eta_1=\eta_2 = 1.9$, $\delta = 0$, $b = 1$, $L = 20.0$.}
\label{fig:Cpm}
\end{figure}

This section presents an asymptotic study of two-soliton interaction in weakly coupled microcavities, governed Eqs.~(\ref{eq:LLE_coupled1})-(\ref{eq:LLE_coupled2}) with small coupling coefficient $\kappa$. When the coupling strength between the cavities is sufficiently weak, the phase difference of the two solitons residing in different microcavities is close the the quantity $\delta$, which is the phase difference of the corresponding injection fields. This phase difference is responsible for either attraction or repulsion of the interacting solitons and for the formation of their bound states.  Let us assume that the coupling coefficient between two cavities and the cavity length mismatch  are small, $\kappa=\epsilon K$ and $V=\epsilon v$ with $\epsilon\ll1$ and look for the solution of Eqs.~(\ref{eq:LLE_coupled1})-(\ref{eq:LLE_coupled2}) in the form
\begin{equation}
A_{1,2}(t,\xi)=i\left[A^{(0)}_{1,2}(\xi)+\epsilon A^{(1)}_{1,2}(t,\xi)+{\cal O}(\epsilon^2)\right]e^{\pm\delta/2},\label{eq:ansatz}
\end{equation}
where $A^{(0)}_{1,2}(\xi)=A^{(0)}[\xi-\xi_{1,2}(\tau)]$ is the soliton solution of the first (second) equation in Eqs.~(\ref{eq:LLE_coupled1})-(\ref{eq:LLE_coupled2}) with $\delta=-\pi$ ($\delta=\pi$) and $\kappa=0$, $\xi_{1,2}(\tau)$ are the slowly varying in time coordinates of the two  solitons 
($\tau=\epsilon t$), and the sign "$+$" ("$-$") corresponds to the first (second) soliton. Substituting Eq. (\ref{eq:ansatz}) into Eqs.~(\ref{eq:LLE_coupled1})-(\ref{eq:LLE_coupled2}) and collecting zero order terms in $\epsilon$ we get for $A^{(0)}_{1,2}$ two uncoupled LLEs   which are satisfied automatically. Collecting first order terms in $\epsilon$ we get
\[{\cal L}_{1,2}A^{(1)}_{1,2}=\psi_{1,2}\left(\partial_\tau \xi_{1,2}\mp \frac{v}{2}\right)+iK A^{(0)}_{2,1}\exp(\mp i\delta),\]
where ${\cal L}_{1,2}$ is the linear operators obtained by linearization of an uncoupled LLE at the unperturbed soliton solutions $A^{(0)}_{1,2}$ and $\psi_{1,2}=\psi(\xi-\xi_{1,2})=\partial_\xi A^{(0)}_{1,2}$ are the neutral modes of these operators corresponding to the translational symmetry of Eqs.~(\ref{eq:LLE_coupled1})-(\ref{eq:LLE_coupled2}). Solvability conditions of these two equations yield the asymptotic equations for the evolution of the coordinates of the two solitons in slow time $\tau$:
\begin{equation}
\partial_\tau \xi_{1,2}=\pm \frac{v}{2}-K\langle \psi^\dagger_{1,2}\cdot iA^{(0)}_{2,1}\exp(\mp i\delta)\rangle,\label{eq:int}
\end{equation} 
where $\psi^\dagger_{1,2}$ are the neutral modes of the adjoint operators, ${\cal L}^\dagger_{1,2}\psi^\dagger(\xi-\xi_{1,2})=0$, satisfying the normalization condition $\langle\psi^\dagger\cdot\psi\rangle=\int_{-L/2}^{L/2}(\Re\psi^\dagger\Re\psi+\Im\psi^\dagger\Im\psi)d\xi=1$. Using the symmetry properties of the soliton solution $A^{(0)}(\xi)=A^{(0)}(-\xi)$ and $\psi^\dagger(\xi)=-\psi^\dagger(-\xi)$, where $\xi=0$ corresponds to the location of the soliton's intensity peak, Eq.~(\ref{eq:int}) can be reduces to the following interaction equations:
\begin{eqnarray}
\partial_\tau \xi_{+}=-K C_+(\xi_-)\sin\delta,\label{eq:int1}\\
\partial_\tau \xi_{-}=-v-K C_-(\xi_-)\cos\delta,\label{eq:int2}
\end{eqnarray}
where $\xi_\pm=\xi_2\pm\xi_1$ and
\begin{eqnarray}
C_+(\xi_-)=\int_{-L/2}^{L/2}\left[\Re A^{(0)}(\xi)\Re\psi^\dagger(\xi-\xi_-)-\Re A^{(0)}(\xi-\xi_-)\Re\psi^\dagger(\xi)\right.\nonumber\\
+\left.\Im A^{(0)}(\xi)\Im\psi^\dagger(\xi-\xi_-)-\Im A^{(0)}(\xi-\xi_-)\Im\psi^\dagger(\xi)\right]d\xi
\end{eqnarray}
\begin{eqnarray}
C_-(\xi_-)=\int_{-L/2}^{L/2}\left[\Im A^{(0)}(\xi)\Re\psi^\dagger(\xi-\xi_-)-\Im A^{(0)}(\xi-\xi_-)\Re\psi^\dagger(\xi)\right.\nonumber\\
-\left.\Re A^{(0)}(\xi)\Im\psi^\dagger(\xi-\xi_-)+\Re A^{(0)}(\xi-\xi_-)\Im\psi^\dagger(\xi)\right]d\xi.
\end{eqnarray}
The dependence of the quantities $C_{+}$ and $C_-$, computed numerically as functions of the fast time $\xi$, is shown in Fig.~\ref{fig:Cpm} by the dotted and solid lines, respectively. 

We first consider the interaction equations~(\ref{eq:int1}) and~(\ref{eq:int2}) in the case where there is no cavity length mismatch, i.e., $v = 0$.  
In this situation, the zeros of the function $C_-$ correspond to the stationary states of Eq.~(\ref{eq:int2}), which represent soliton states with fixed separation $\xi_-$.  
In Fig.~\ref{fig:Cpm}, these zeros are located at $\xi = 0$, $\xi = \pm \xi_* \approx \pm 4.34572$, and $\xi = \pm L/2$, where the points $\xi = \pm L/2$ correspond to a single state due to the periodic boundary conditions.  
Since $C_+ = 0$ at $\xi = 0$ and $\xi = \pm L/2$, the corresponding soliton clusters are stationary along the $\xi$-axis. Below they will be referred as soliton `bunching' and `anti-bunching,' respectively.  
In contrast, the soliton states with fixed separations $\xi = \pm \xi_*$ move in opposite directions along $\xi$-axis when $\delta \neq 0, \pm \pi$, since for these states $C_+(\xi) \neq 0$.  
Note that the characteristic distance $\xi_*$ depends on the soliton size and is almost independent of the cavity round-trip length $L$.
Therefore, such a moving soliton cluster can be regarded as a classical soliton bound state, arising from the interaction of solitons through their tails.

\begin{figure}[htbp]
  \centering
  \begin{minipage}{0.48\textwidth}
    \centering
    \includegraphics[width=\linewidth]{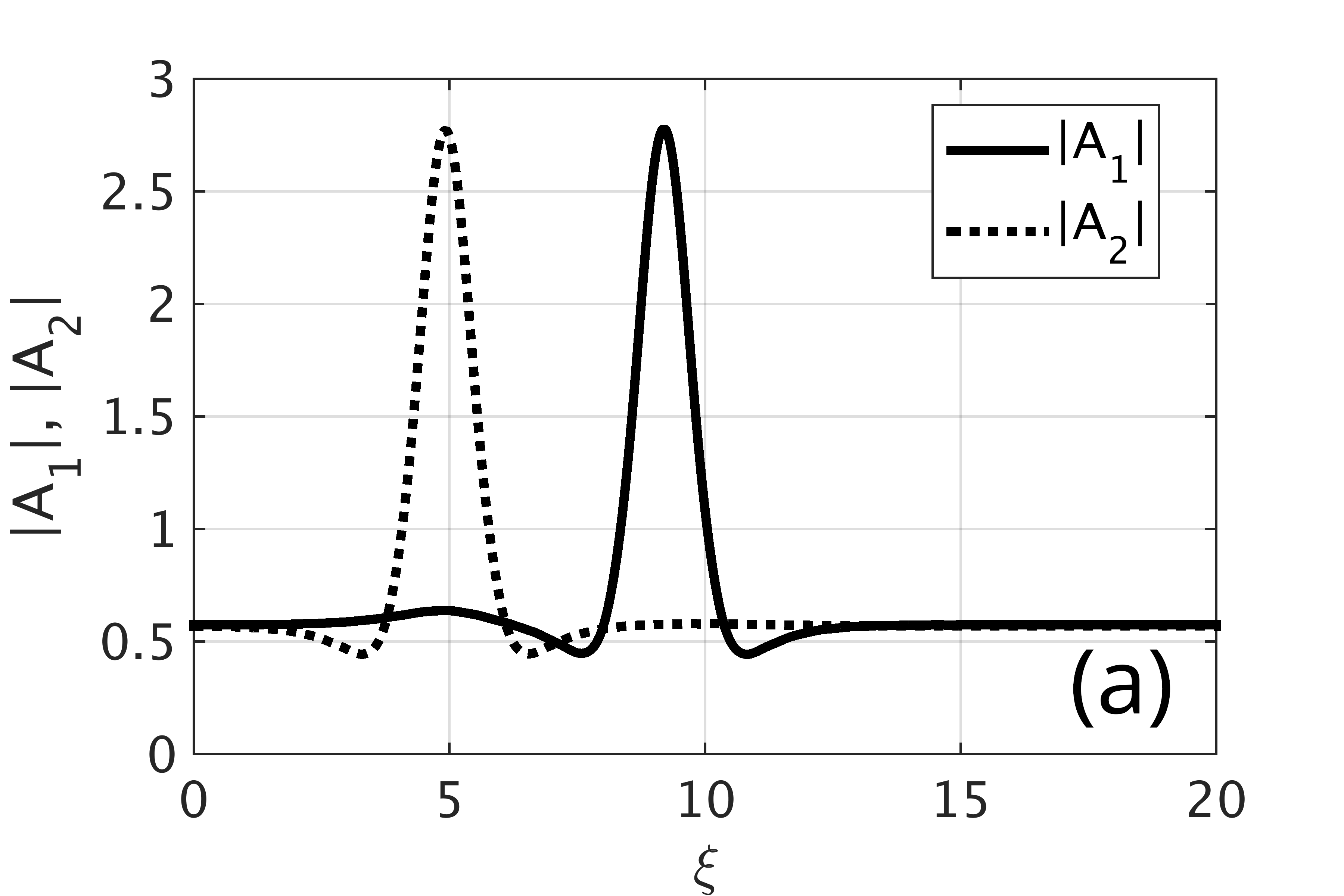}
  \end{minipage}
  \hfill
  \begin{minipage}{0.48\textwidth}
    \centering
    \includegraphics[width=\linewidth]{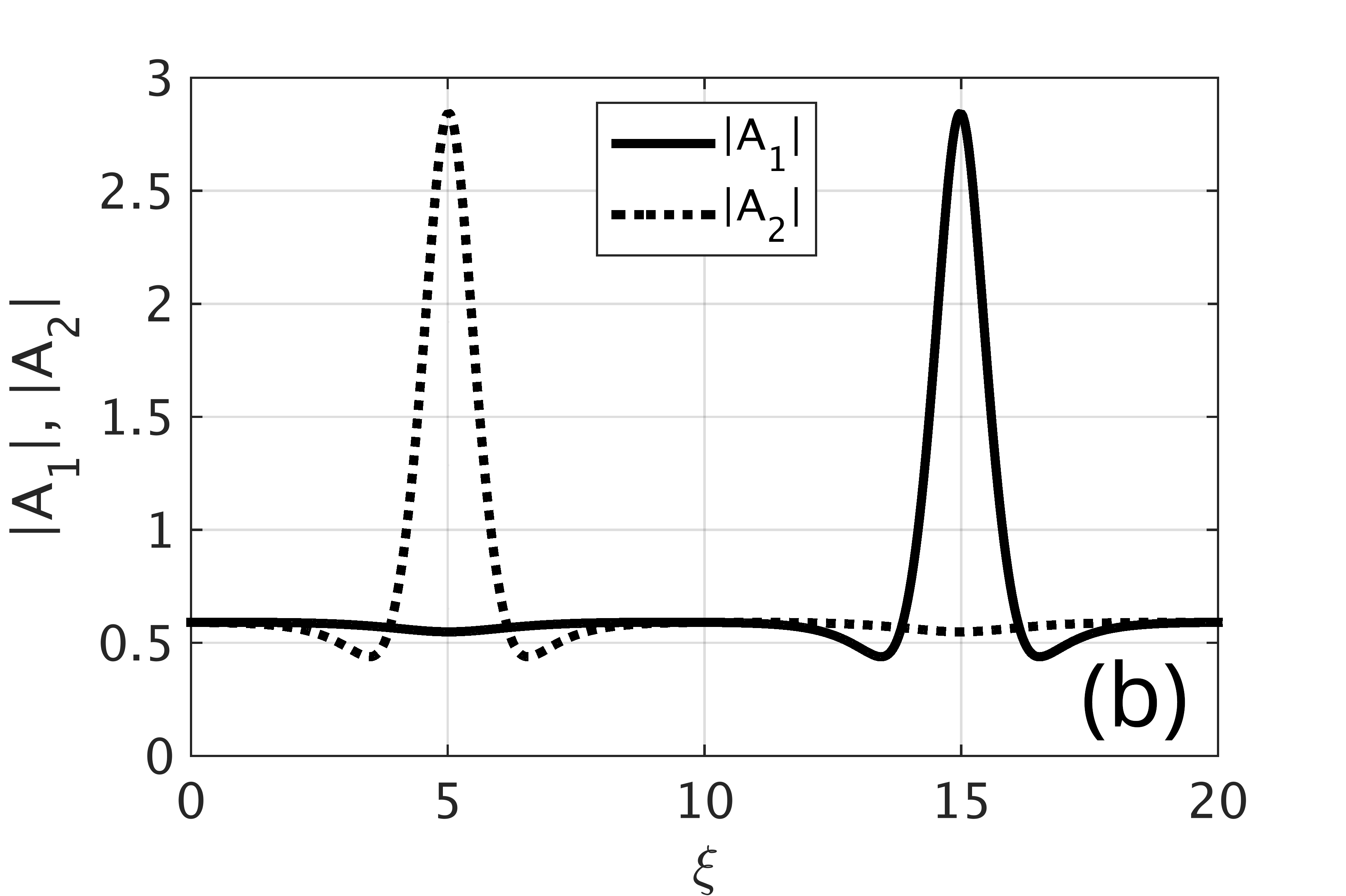}
  \end{minipage}
  \vspace{0.2cm}

  \begin{minipage}{0.48\textwidth}
    \centering
    \includegraphics[width=\linewidth]{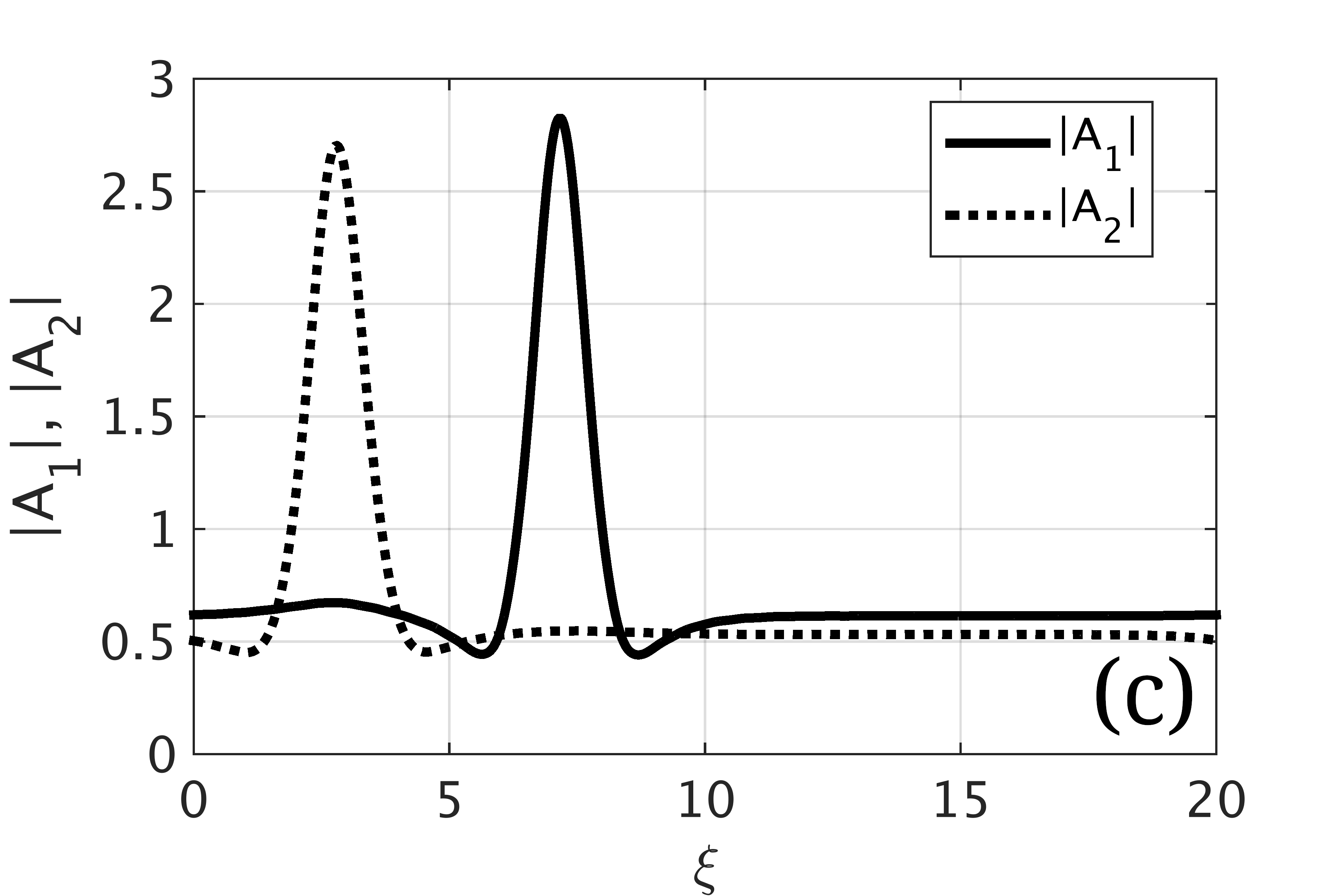}
  \end{minipage}
  \hfill
  \begin{minipage}{0.48\textwidth}
    \centering
    \includegraphics[width=\linewidth]{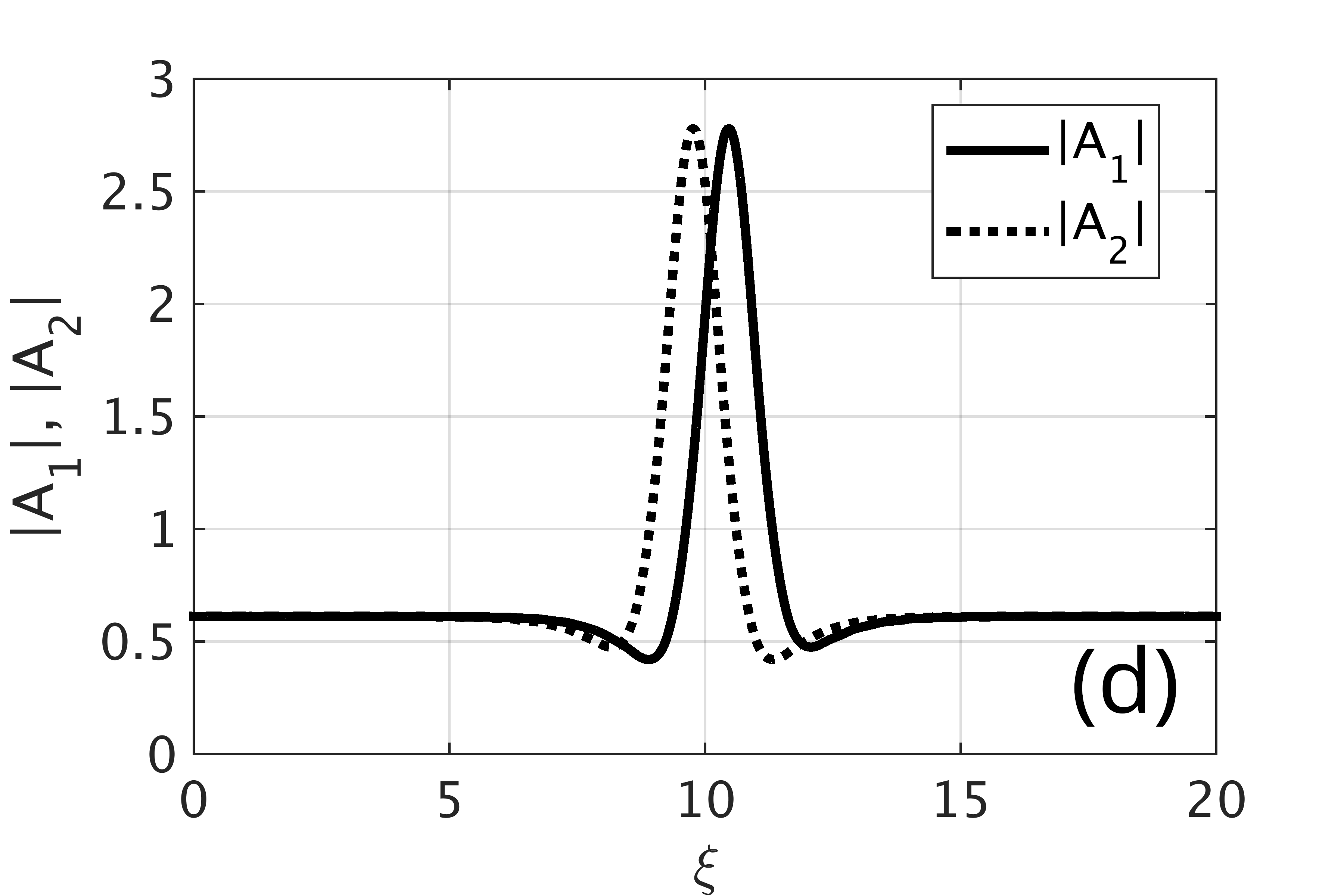}
  \end{minipage}
\caption{Amplitude of the first and second solitons, $|A_1|$ and $|A_2|$, corresponding to stable states obtained via numerical integration of the coupled LLEs.  
(a) Bound state of moving solitons at $\delta = 2.64$, $V = 0$, $\eta_1=\eta_2=1.9$, and $L=20.0$.  
(b) Soliton anti-bunching at $\delta = 0$, $V = 0$, $\eta_1=\eta_2=1.9$, and $L=12.0$.  
(c) Same as (a) but with different injection rates $\eta_1=2.0$ and $\eta_2=1.8$. 
(d) Soliton bunching at $\delta = 0$, $V = 0.2$, and $L=20.0$.
Other parameters: $\theta = 3.5$, $\eta_1=\eta_2 = 1.9$, $b = 1$.}
  \label{fig:4panel}
\end{figure}

Since, as can be seen from Fig.~\ref{fig:Cpm}, the derivative $\partial_{\xi_-}C_{-}(\xi)$ is negative at $\xi = 0, \pm L/2$, it follows from Eqs.~(\ref{eq:int1}) and~(\ref{eq:int2}) that the corresponding bunching and anti-bunching states are linearly stable for $-\pi/2 < \delta < \pi/2$ and unstable for $\pi/2 < |\delta| < \pi$. In contrast, the `moving' bound states located at $\xi = \pm \xi_*$ are linearly stable for $\pi/2 < |\delta| < \pi$ and unstable within the interval $-\pi/2 < \delta < \pi/2$. Moving bound states obtained by numerical integration of Eqs.~(\ref{eq:LLE_coupled1})-(\ref{eq:LLE_coupled2}) for $\eta_1=\eta_2$ and $\eta_1>\eta_2$ are shown in Fig.~\ref{fig:4panel}(a) and (c), respectively. Note, that since the anti-bunching state illustrated in Fig.~\ref{fig:4panel}(b) corresponds to the maximum possible soliton separation, $\xi_-=L/2$, and thus the weakest soliton overlap, its stability is intrinsically fragile and can be easily disrupted by even very small perturbations, such as a slight cavity length mismatch.

At the critical value $\delta = \pm \pi/2$, the right-hand side of Eq.~(\ref{eq:int2}) with $v=0$ vanishes, implying that, to first order, the entire interval $-L/2 \leq \xi \leq L/2$ is filled with stationary bound states. To resolve this degeneracy, one would need to include higher-order terms in the interaction equations. However, a detailed analytical treatment of this case lies beyond the scope of the present work. Instead, in the following sections, we numerically investigate the behavior of Eqs.~(\ref{eq:LLE_coupled1})-(\ref{eq:LLE_coupled2}) both near and away from the point $\delta = \pm \pi/2$.

When the cavity length mismatch is nonzero, $V \neq 0$ in Eqs.~(\ref{eq:LLE_coupled1})-(\ref{eq:LLE_coupled2}), and sufficiently large, soliton clusters with fixed separations are destroyed, resulting in solitons traveling at different velocities. However, Eq.~(\ref{eq:int2}) shows that if $v$ is sufficiently small and $\cos\delta \neq 0$, these clusters can persist. In particular, for a fixed phase difference within the interval $-\pi/2 < \delta < \pi/2$, the bunching state develops a nonzero soliton separation $\xi_-$ as $|v|$ increases, as illustrated in Fig.~\ref{fig:4panel}(d). This state disappears at a saddle-node bifurcation when $|v| > K C_-^{\max}(\xi_-)$, where $C_-^{\max}$ denotes the maximum of the function $C_-$ (solid line in Fig.~\ref{fig:Cpm}). The corresponding maximum soliton separation in the bunching state, $\xi_-^{\max}$, equals the absolute value of the abscissa at which this maximum occurs.  
This behavior is characteristic of a synchronization–desynchronization transition, in which increasing detuning parameters lead to the breakdown of stable phase-locked states. Furthermore, for $v \neq 0$ and $\delta \neq 0, \pm \pi/2, \pm \pi$, the drift velocity $\partial_\tau \xi_+$ of the bunching (or anti-bunching) states is nonzero, because for $v \neq 0$, Eq.~(\ref{eq:int2}) implies that the soliton separation is not zero ($L/2$), and hence $C_+ \neq 0$ in Eq.~(\ref{eq:int1}).

Note that the appearance of the cosine and sine functions of the phase difference in the equations governing soliton separation and center-of-mass velocity closely resembles earlier results on dissipative soliton interactions in Ginzburg--Landau-type models; see, e.g., \cite{vladimirov2001stable,turaev2012long}. However, in the present case, there are no separate evolution equations for the individual soliton phases, as these phases are fixed by the external injection. Another characteristic property of the interaction equations~(\ref{eq:int1}) and~(\ref{eq:int2}) is that they were derived under the assumption of weak coupling between the cavities and are valid for arbitrarily small soliton separations $\xi_-$.

\section{Numerical results}
\subsection{Continuous wave solutions}
\begin{figure}[thb]
\centering
\includegraphics[width=0.9\linewidth]{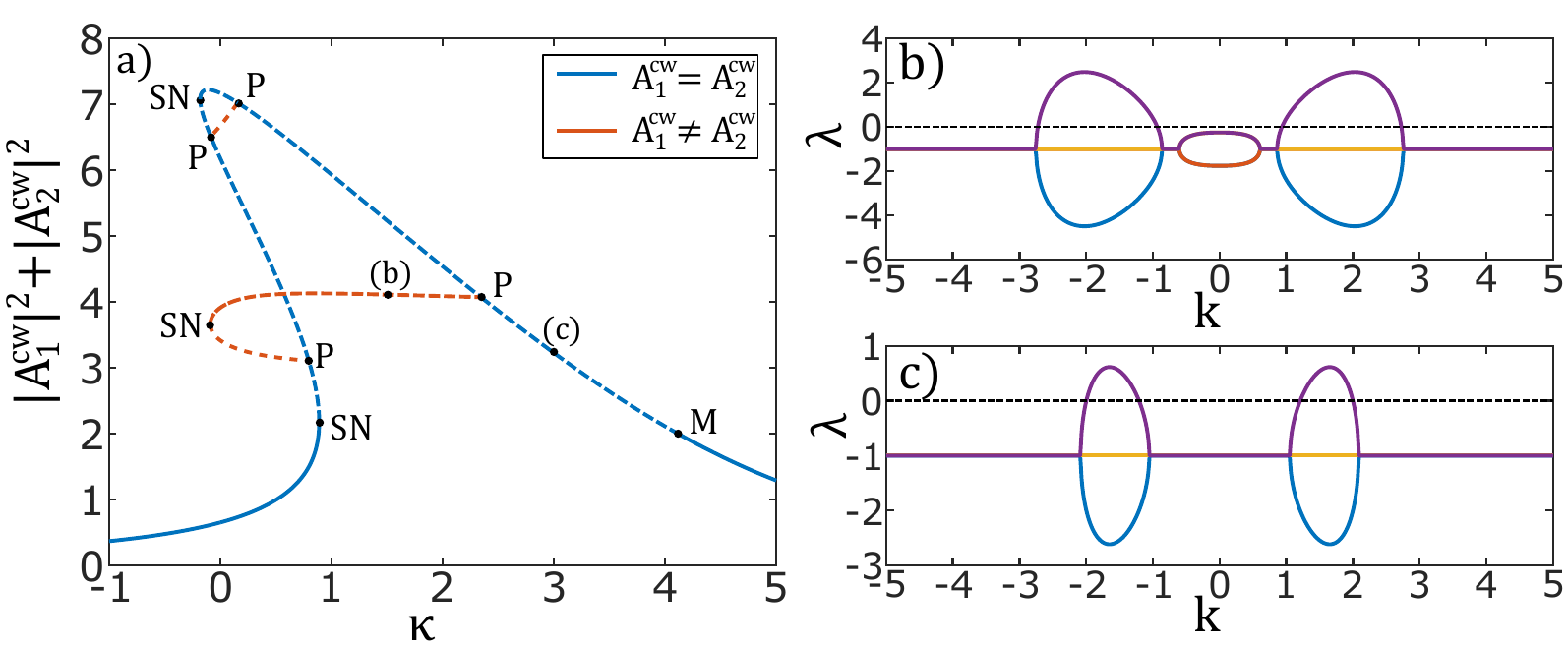}
\caption{
a) Total intensity ($|A_1|^2 + |A_2|^2$) of CW solutions as a function of the coupling coefficient $\kappa$. Symmetric solutions ($|A_1|^2 = |A_2|^2$) are shown in blue, and asymmetric solutions ($|A_1|^2 \neq |A_2|^2$) in red. Solid lines represent stable states; dashed lines indicate unstable ones. The points labeled `P' and `SN' mark the pitchfork and saddle-node bifurcations, respectively. `M' denotes the onset of modulational instability.  
b) Real parts of the dominant eigenvalues from the linear stability spectrum of a modulationally unstable asymmetric state at $\kappa = 1.5$.  
c) Same as in (b), but for a modulationally unstable symmetric state at $\kappa = 3$.  
Parameters: $\theta = 3.5$, $\eta_1=\eta_2 = 1.9$, $V = 0$, $\delta = 0$, and $b = 1$.
}
\label{fig:uniform}
\end{figure}

First, we analyze continuous wave (CW) solutions by setting the temporal and spatial derivatives in Eqs.~(\ref{eq:LLE_coupled1})-(\ref{eq:LLE_coupled2}) to zero. Figure~\ref{fig:uniform}(a) illustrates how the total CW intensity, $|A_1|^2 + |A_2|^2$, varies with the coupling coefficient $\kappa$ in the anomalous dispersion regime ($b=1$), under zero pumping phase difference and drift ($\delta = V = 0$). It is seen from the figure that resulting curve exhibits the characteristic nonlinear resonance shape.

\begin{figure}[htb]
\centering
\includegraphics[width=0.45\linewidth]{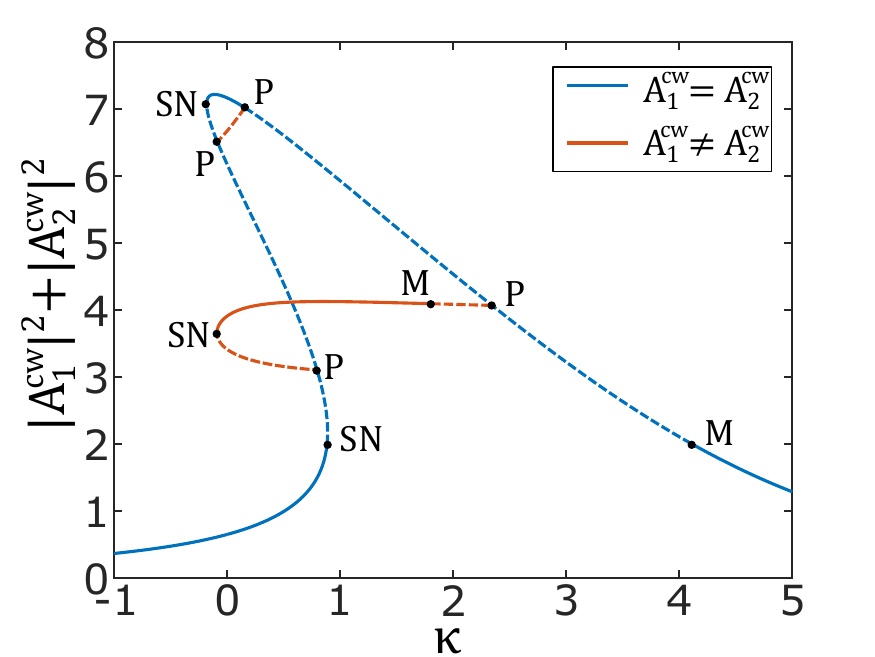}
\caption{
a) Total intensity ($|A_1|^2 + |A_2|^2$) of CW solutions as a function of the coupling coefficient $\kappa$ for the case of normal dispersion. Parameters are the same as in Fig.~\ref{fig:uniform}(a), except with $b = -1$.
}

\label{fig:normal_disp_uni}
\end{figure}

Two distinct types of CW solutions are identified in Fig.~\ref{fig:uniform}(a): symmetric solutions, where the intensities in both cavities are equal ($|A_1|^2 = |A_2|^2$), and asymmetric ones with unequal intensities ($|A_1|^2 \neq |A_2|^2$). The latter emerge through a pitchfork bifurcation and are always unstable in the anomalous dispersion regime ($b=1$). Note that
due to the symmetry of the model equations (\ref{eq:LLE_coupled1})-(\ref{eq:LLE_coupled2}) with $\delta=V=0$ and $\eta_1=\eta_2$ there are two asymmetric solutions with the same total intensity, one with $|A_1|^2>|A_2|^2$ and another with $|A_1|^2<|A_2|^2$. To check the stability we introduced perturbations in the form $A_{1,2}= A_{1,2} + \epsilon_{1,2}e^{\lambda t + i k \xi}$. The upper asymmetric states (characterized by higher total intensity) are unstable with respect to perturbations having wavenumber $k = 0$. The lower asymmetric states are modulationary unstable, see Fig.~\ref{fig:uniform}(b). The lower branch of the symmetric state is stable up to $\kappa_{SN} \approx 0.89$ where a saddle-node bifurcation takes place; the upper branch is modulationally unstable up to $\kappa \approx 4.1$, then it stabilizes. Figure~\ref{fig:normal_disp_uni} is similar to Fig.~\ref{fig:uniform}(a), but was calculated for the normal dispersion case, $b = -1$. It is seen that in this case a small part of the upper symmetric state as well as a considerable part of the lower asymmetric states are stable. Furthermore, within a certain interval of coupling coefficients $\kappa$ the system exhibits multistability between two symmetric and a pair of lower asymmetric states. 

\subsection{Effect of pump phase difference on soliton interaction}

\begin{figure}[th]
\centering
\includegraphics[width=0.5\linewidth]{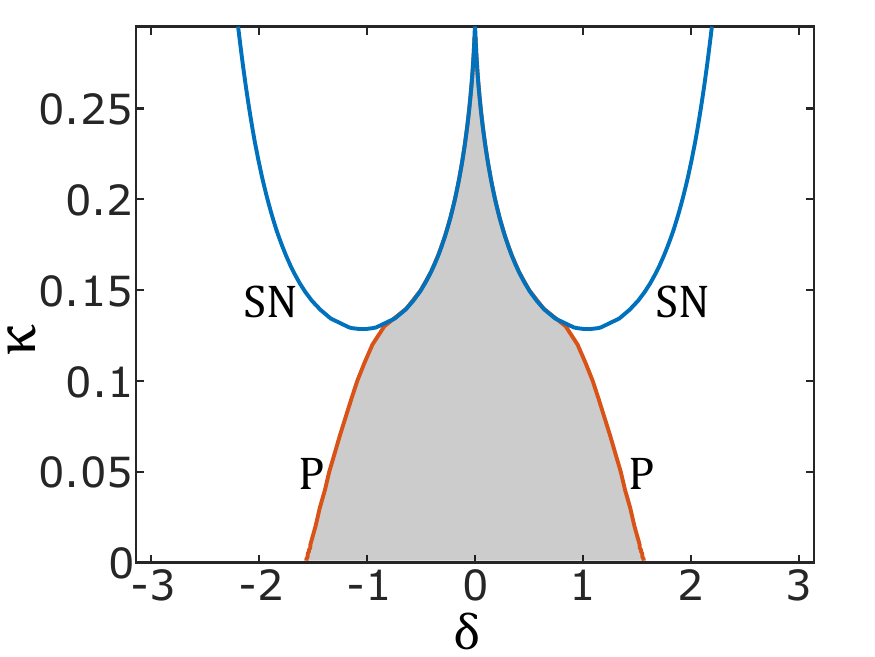}
\caption{Shaded area indicates the parameter region where bunching soliton clusters are stable. Parameters: $\theta = 3.5$, $\eta_1=\eta_2 = 1.9$, $V = 0$, $b = 1$. }
\label{fig:kappa_del}
\end{figure}

\begin{figure}[tbh]
\centering
\includegraphics[width=0.9\linewidth]{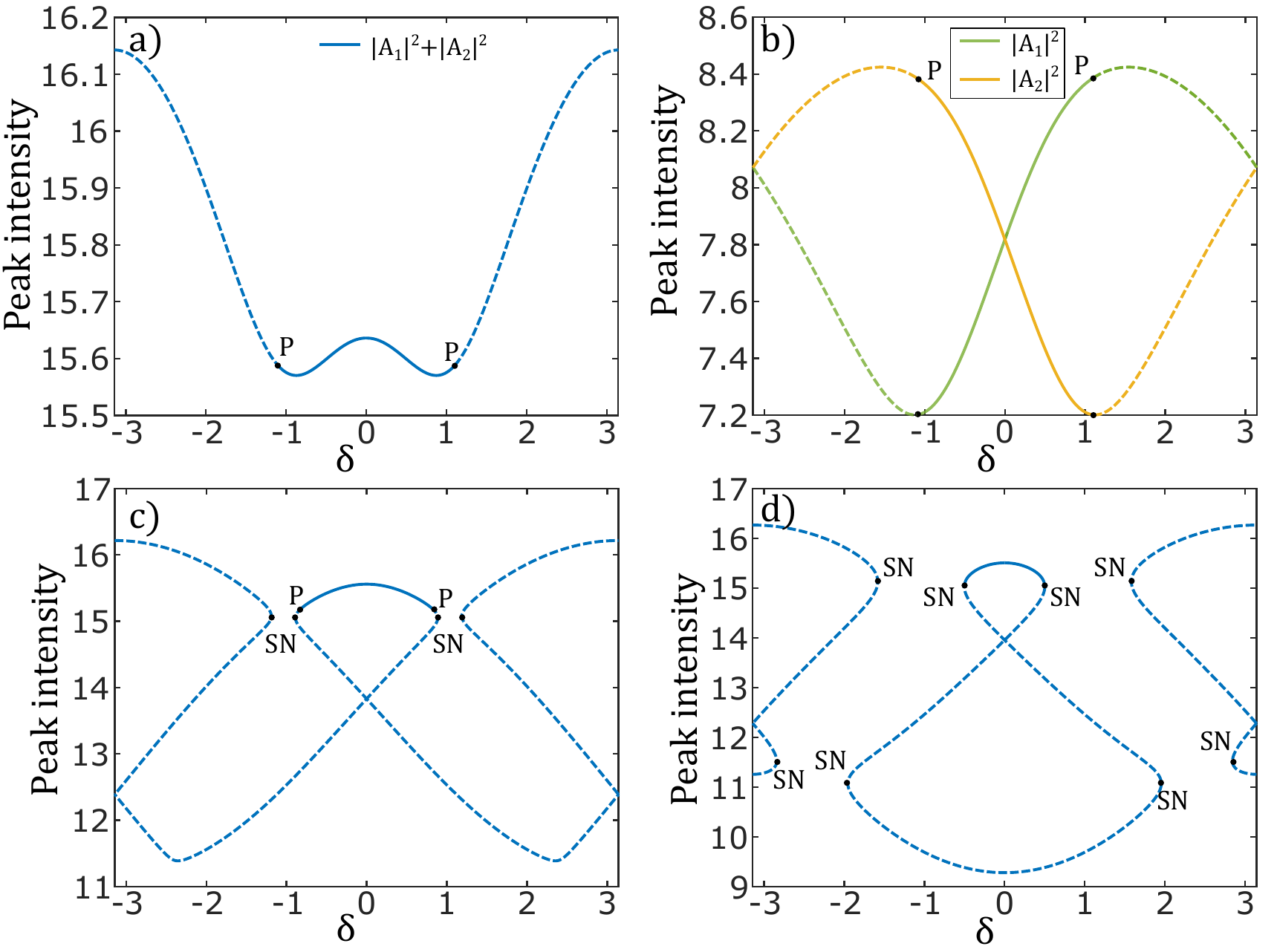}
\caption{Soliton peak intensities as a function of $\delta$. 
(a) Sum of the soliton peak intensities, $|A_1|^2 + |A_2|^2$, for $\kappa = 0.1$. 
(b) Individual peak intensities, $|A_1|^2$ and $|A_2|^2$, for $\kappa = 0.1$. 
(c) Sum of the soliton peak intensities for $\kappa = 0.13$. 
(d) Sum of the soliton peak intensities for $\kappa = 0.15$. 
Parameters: $\theta = 3.5$, $\eta_1 = \eta_2 = 1.9$, $V = 0$, and $L = 50$.}
\label{fig:intensity_delta}
\end{figure}

In this section, we present numerical calculation of the branches of soliton clusters as functions of the pump phase difference $\delta$, assuming zero cavity length mismatch and equal injection rates,  $V = 0$ and $\eta_1=\eta_2$. 
The gray area in Fig.~\ref{fig:kappa_del} indicates the domain of existence, in the $\delta$–$\kappa$ plane, of a stable bunching state, characterized by two solitons occupying identical positions on the $\xi$-axis. This region is bounded by a pitchfork (P) and a saddle-node  (SN) bifurcations. 
Figure~\ref{fig:intensity_delta} shows the dependence of the soliton peak intensities, $|A_1|^2$ and $|A_2|^2$, as well as their sum, on the phase difference $\delta$, computed for various values of the coupling strength $\kappa$. Stable and unstable solutions are shown with solid and dashed lines, respectively. 
For small values of $\kappa$, the stability range of the bunching state closely matches the interval $\pi < \delta < \pi$, in agreement with our asymptotic analysis. As the coupling strength increases, this stability range narrows and eventually disappears. As it is seen from Fig.~\ref{fig:intensity_delta}(b) for $\delta\neq0$ there are two symmetric bunching states with different soliton peak intensities.

\begin{figure}[ht]
\centering
\includegraphics[width=0.7\linewidth]{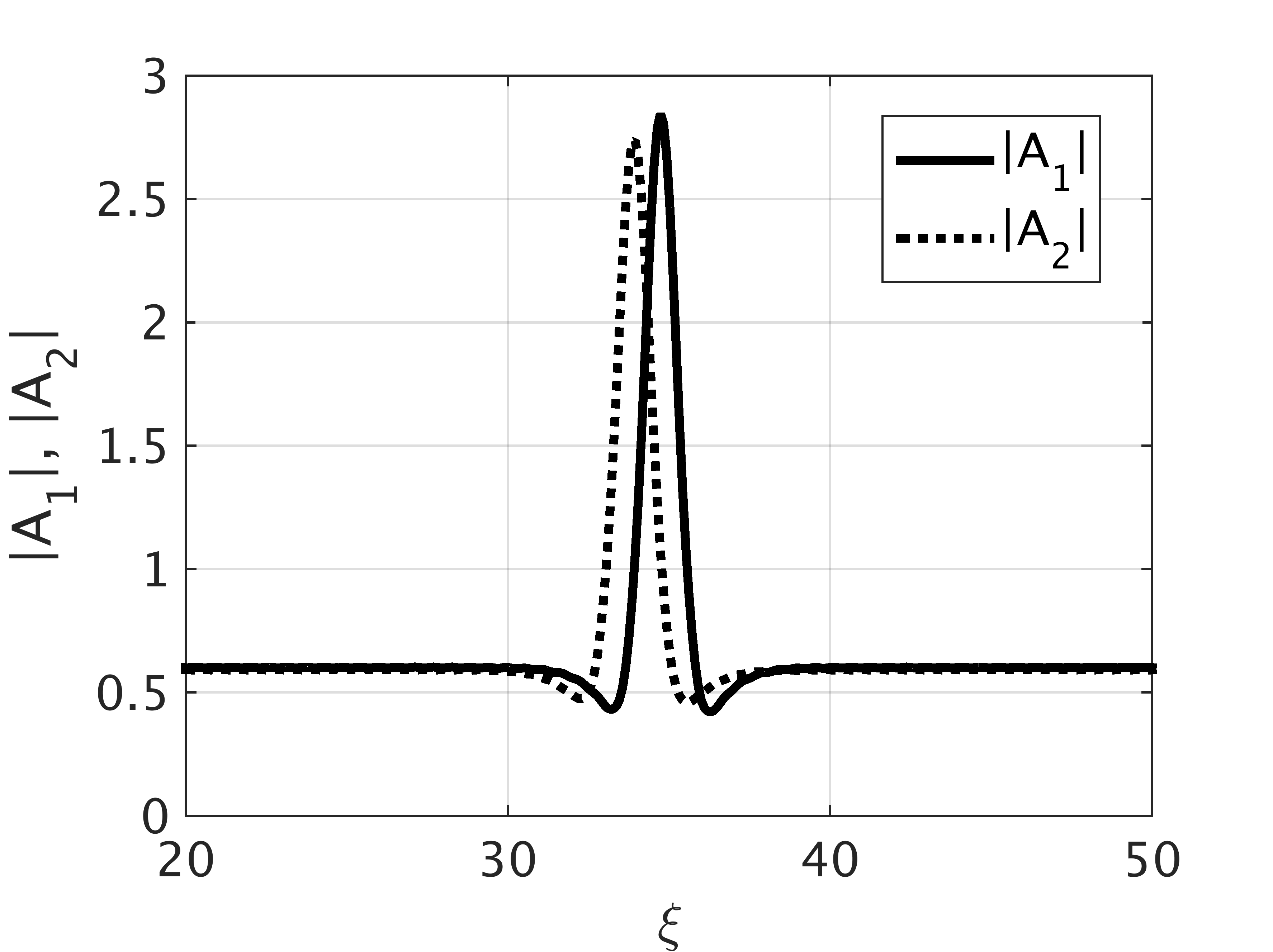}
\caption{One of the two stable soliton clusters bifurcating from the soliton bunching state at the pitchfork bifurcation point P.  $\kappa = 0.1$. Other parameters are the same as in Fig.~\ref{fig:intensity_delta}.\label{fig:moving_p}
}
\end{figure}

\begin{figure}[ht]
\centering
\includegraphics[width=0.9\linewidth]{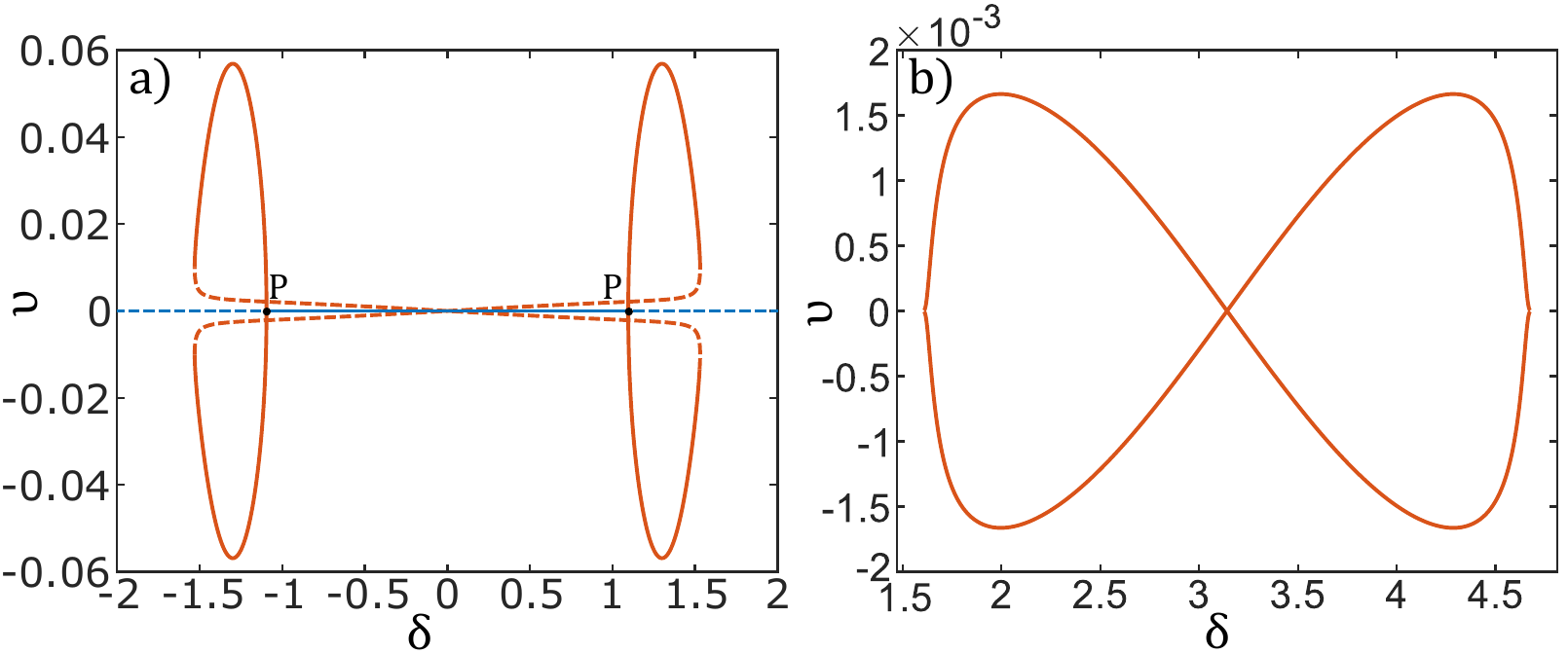}
\caption{
(a) Velocity dependence of moving bound states (red curve) emerging via a pitchfork bifurcation from the symmetric stationary state (blue curve).  
(b) Velocity dependence of the moving soliton bound states shown in Fig.~\ref{fig:4panel}(a) for $\kappa = 0.1$. Other parameters are as in Fig.~\ref{fig:intensity_delta}.
}
\label{fig:moving_del}
\end{figure}

The pitchfork bifurcations labeled `P' in Fig.~\ref{fig:kappa_del} and Fig.~\ref{fig:intensity_delta}(a,b) generate two clusters of soliton solutions propagating in opposite directions. As it is illustrated by Fig.~\ref{fig:moving_p}, these clusters are characterized by slightly different soliton peak powers and a finite soliton separation, which increases with the distance from the bifurcation point P. The dependence of their velocity on the pump phase difference $\delta$ is shown in Fig.~\ref{fig:moving_del}(a). Their stability ranges are relatively narrow and confined to neighborhoods of the pitchfork bifurcation point. As the coupling strength $\kappa$ tends to zero, these stability intervals contract toward the critical value $\delta = \pm \pi/2$. In contrast, the unstable portions of the corresponding moving branches persist in this limit and coincide with the unstable soliton bound states that our asymptotic analysis predicts to exist for $-\pi/2 < \delta < \pi/2$.

Figure~\ref{fig:moving_del}(b) shows the velocity dependence of the stable soliton bound state depicted in Fig.~\ref{fig:4panel}(a), plotted as a function of the pump phase difference $\delta$. The existence range of this stable solution lies close to the interval $\pi/2 < \delta < 3\pi/2$, in agreement with the prediction for the small coupling limit, $\kappa \to 0$. The small velocities observed for the soliton bound states are attributed to weak soliton interactions, which result from sufficiently large separation between individual solitons within the bound state. 

Note that for $V=0$, the branch of stable soliton bound states shown in Fig.~\ref{fig:moving_del}(b), characterized by an almost constant and large soliton separation $\xi_- \approx \xi_*$, is disconnected from the bunching-state solution branch at $\xi_- = 0$. 
By contrast, the unstable segments of the moving soliton state, shown by the red dotted line in Fig.~\ref{fig:moving_del}(a), correspond to unstable soliton bound states predicted by our asymptotic analysis and are connected to the bunching state.





\subsection{Soliton clusters under cavity length mismatch}

\begin{figure}[th]
\centering
\includegraphics[width=0.45\linewidth]{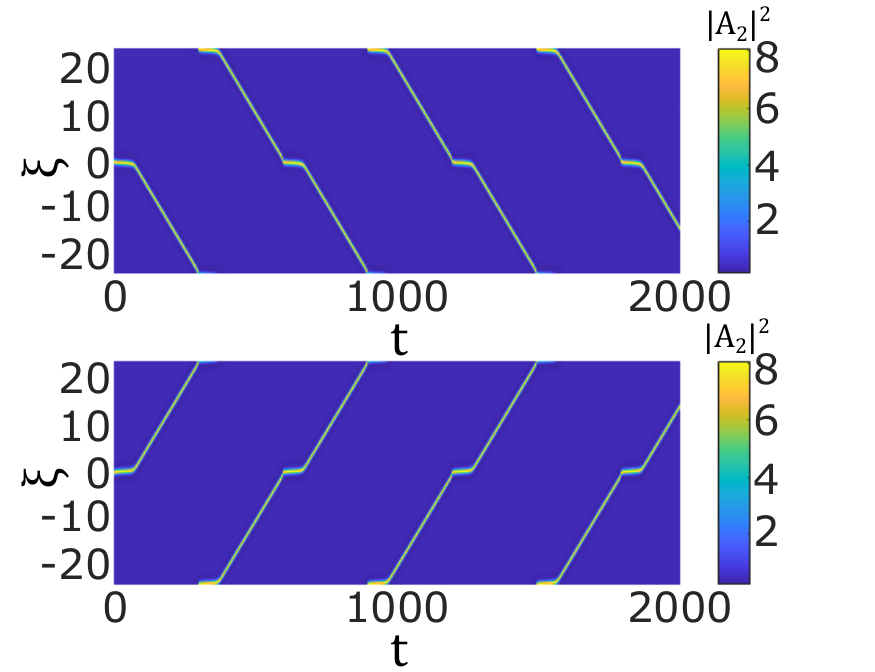}
\caption{Drifting solitons outside the Arnold synchronization tongue. Parameter values: $\theta = 3.5$, $\eta_1=\eta_2 = 1.9$, $\delta = 0$, $b = 1$, $L=50$, and $\kappa=0.1$.}
\label{fig:SN_dist}
\end{figure}

Small cavity length mismatch can be incorporated via a nonzero parameter $V$ in Eqs.~(\ref{eq:LLE_coupled1})-(\ref{eq:LLE_coupled2}). When $V$ is nonzero but remains sufficiently small, solitons in the bunching state develop a finite separation $\xi_-$, which increases with $V$, see Fig.\ref{fig:4panel}(d). However, once $V$ exceeds a critical threshold, the bunching state is destroyed, and the solitons begin to drift apart in opposite directions, as illustrated in Fig.~\ref{fig:SN_dist}.

\begin{figure}[th]
\centering
\includegraphics[width=0.99\linewidth]{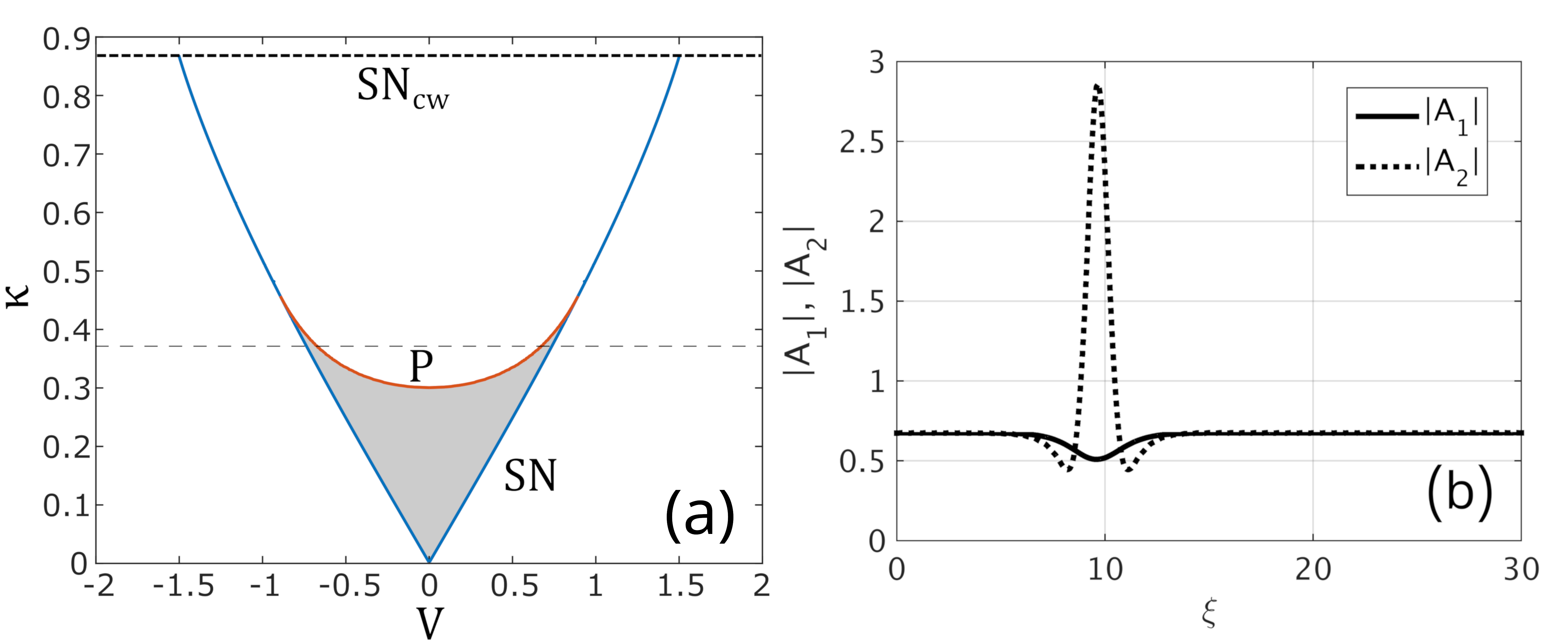}
\caption{
(a) Shaded area indicates the existence domain of stable soliton bunching states, bounded by saddle-node (SN) and pitchfork (P) bifurcations. `SN$_{cw}$' denotes the saddle-node bifurcation of the CW background.  
(b) Drifting single soliton solution above the pitchfork bifurcation threshold, with $V=0.2$ and $\kappa=0.335$. Other parameters are as in Fig.~\ref{fig:SN_dist}.
}

\label{fig:kappa_V}
\end{figure}

The corresponding bifurcation diagram in the $(V$–$\kappa)$ plane, obtained from numerical solutions of Eqs.~(\ref{eq:LLE_coupled1})-(\ref{eq:LLE_coupled2}), is shown in Fig.~\ref{fig:kappa_V}. This diagram exhibits the characteristic structure of a $1:1$ Arnold tongue \cite{arnol1963small,pikovsky2001universal}. The bunching state exists near $V = 0$ and is bounded by two saddle-node bifurcation curves SN, shown as blue lines in Fig.~\ref{fig:kappa_V}(a). When crossing these lines, the stationary stable bunching state merges with an unstable soliton cluster, and a transition to a pair of oppositely drifting solitons takes place; see Fig.~\ref{fig:SN_dist}. Note that stable soliton solitons cannot exist beyond the saddle-node bifurcation of the CW background, marked by the horizontal line $SN_{cw}$.

The stable bunching state occupies the shaded region below the pitchfork bifurcation curve $P$. Upon crossing this curve from below, numerical integration of Eqs.~(\ref{eq:LLE_coupled1})-(\ref{eq:LLE_coupled2}) reveals a sharp transition to one of two bistable states: a soliton in one component and a slightly perturbed continuous wave in the other. An example of such a stable solution, directly related to the bistability of the soliton and its CW background in a solitary LLE, is shown in Fig.~\ref{fig:kappa_V}(b).

\begin{figure}[th]
\centering
\includegraphics[width=0.9\linewidth]{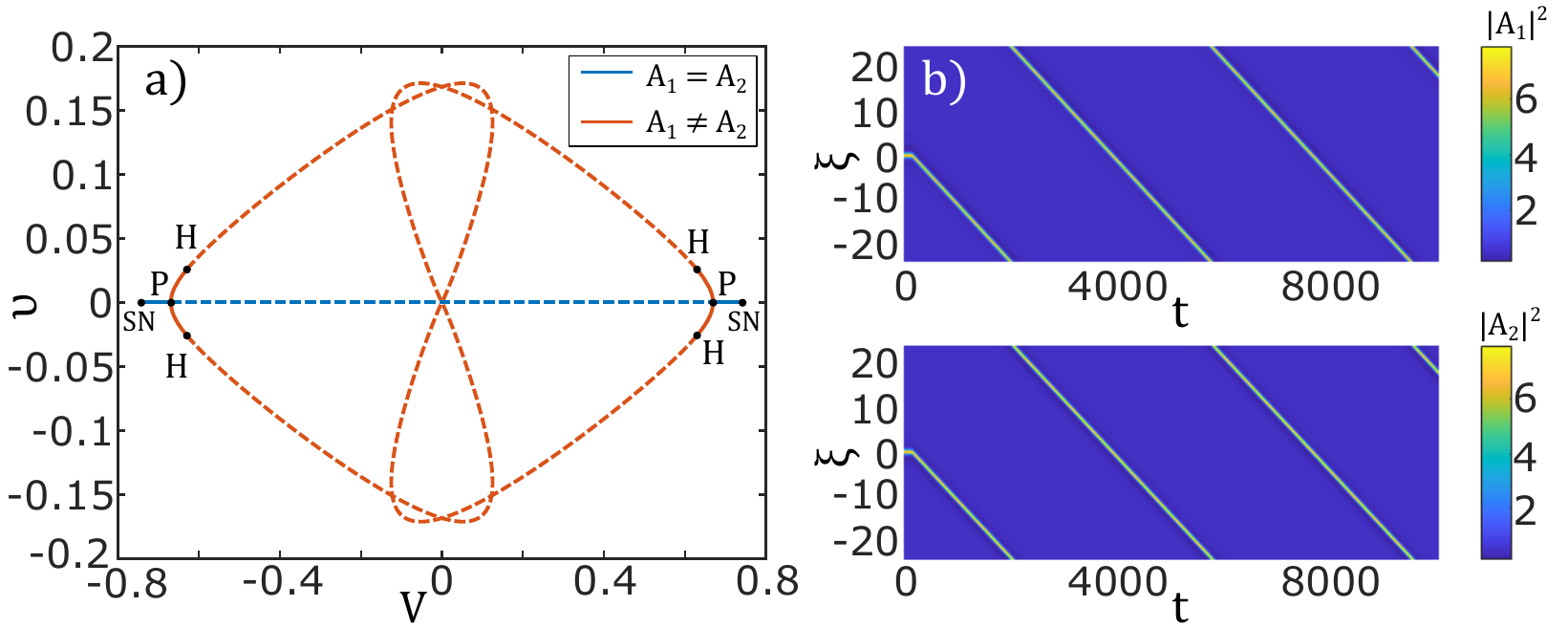}
\caption{
(a) Velocity dependence of moving soliton clusters (red curve) emerging via a pitchfork bifurcation from the symmetric stationary state (blue). Parameters are the same as in Fig.~\ref{fig:kappa_V}(a), with $\kappa = 0.37$.  
(b) Emergence of a stable moving state at $V = 0.65$.
}

\label{fig:moving_V}
\end{figure}

As follows from Fig.~\ref{fig:moving_V}(a), which illustrates the dependence of the soliton cluster velocity $\upsilon$ on $V$, the pitchfork bifurcation is supercritical at sufficiently large $\kappa$. As a result, the destabilization of the synchronized state gives rise to a pair of new solutions. These solutions remain bounded states of two solitons, but they propagate with a nonzero velocity $\upsilon$. Close to the bifurcation point `P', such states are stable; however, they are eventually destabilized through an  Andronov–Hopf bifurcation.

\begin{figure}[htbp]
  \centering
  \begin{minipage}{0.48\textwidth}
    \centering
    \includegraphics[width=\linewidth]{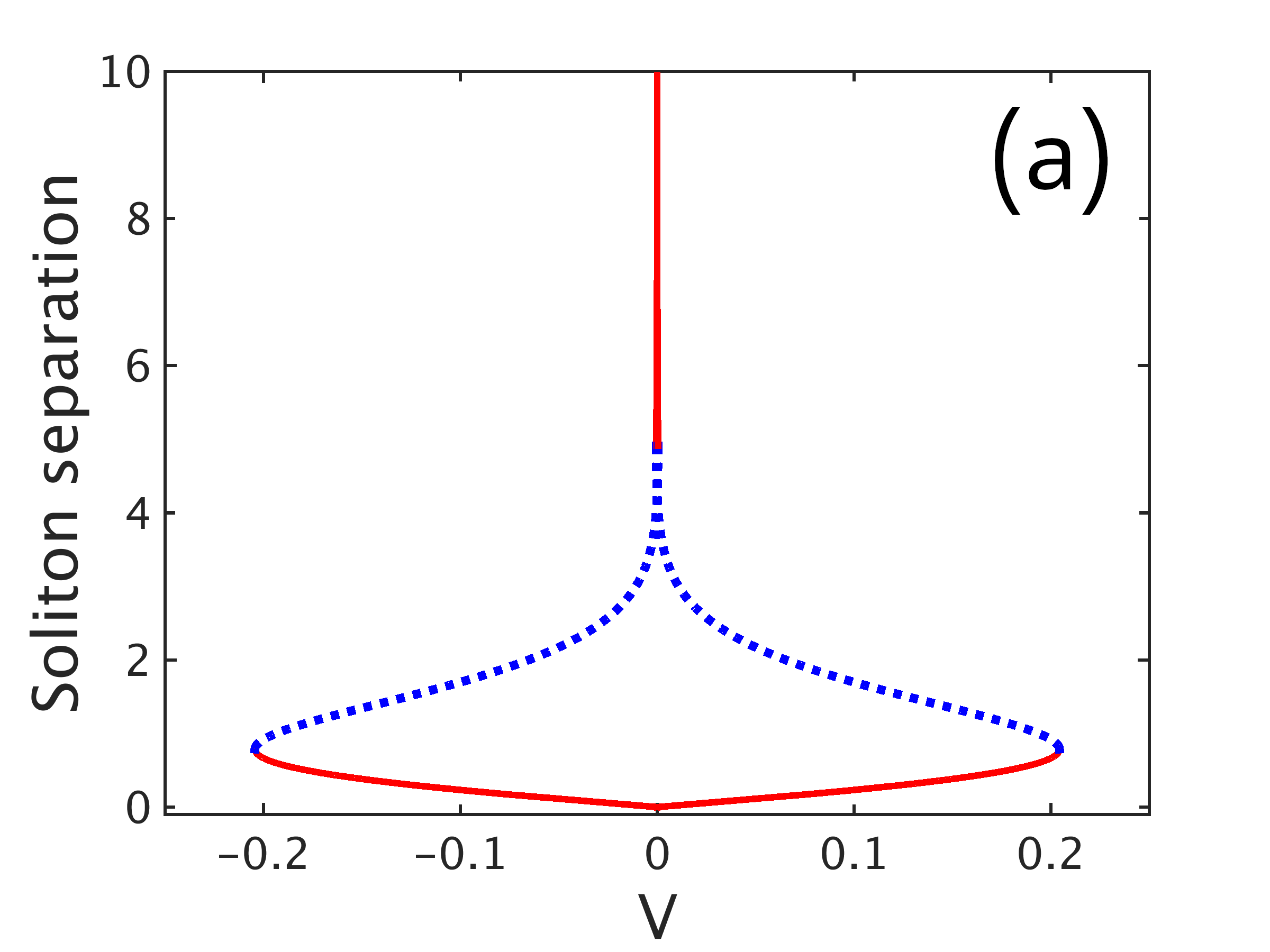}
  \end{minipage}
  \hfill
  \begin{minipage}{0.48\textwidth}
    \centering
    \includegraphics[width=\linewidth]{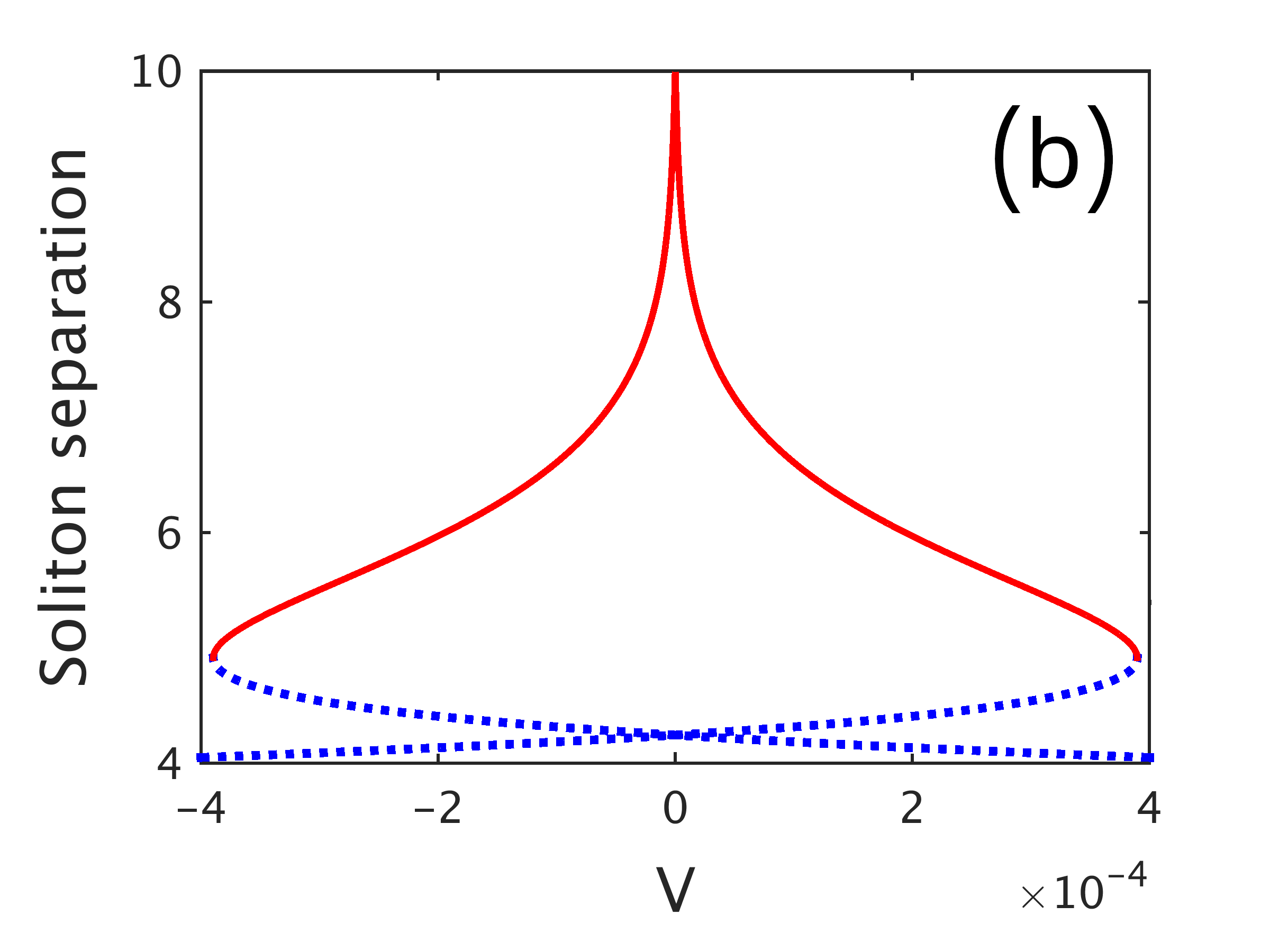}
  \end{minipage}
  \vspace{0.2cm}

  \begin{minipage}{0.48\textwidth}
    \centering
    \includegraphics[width=\linewidth]{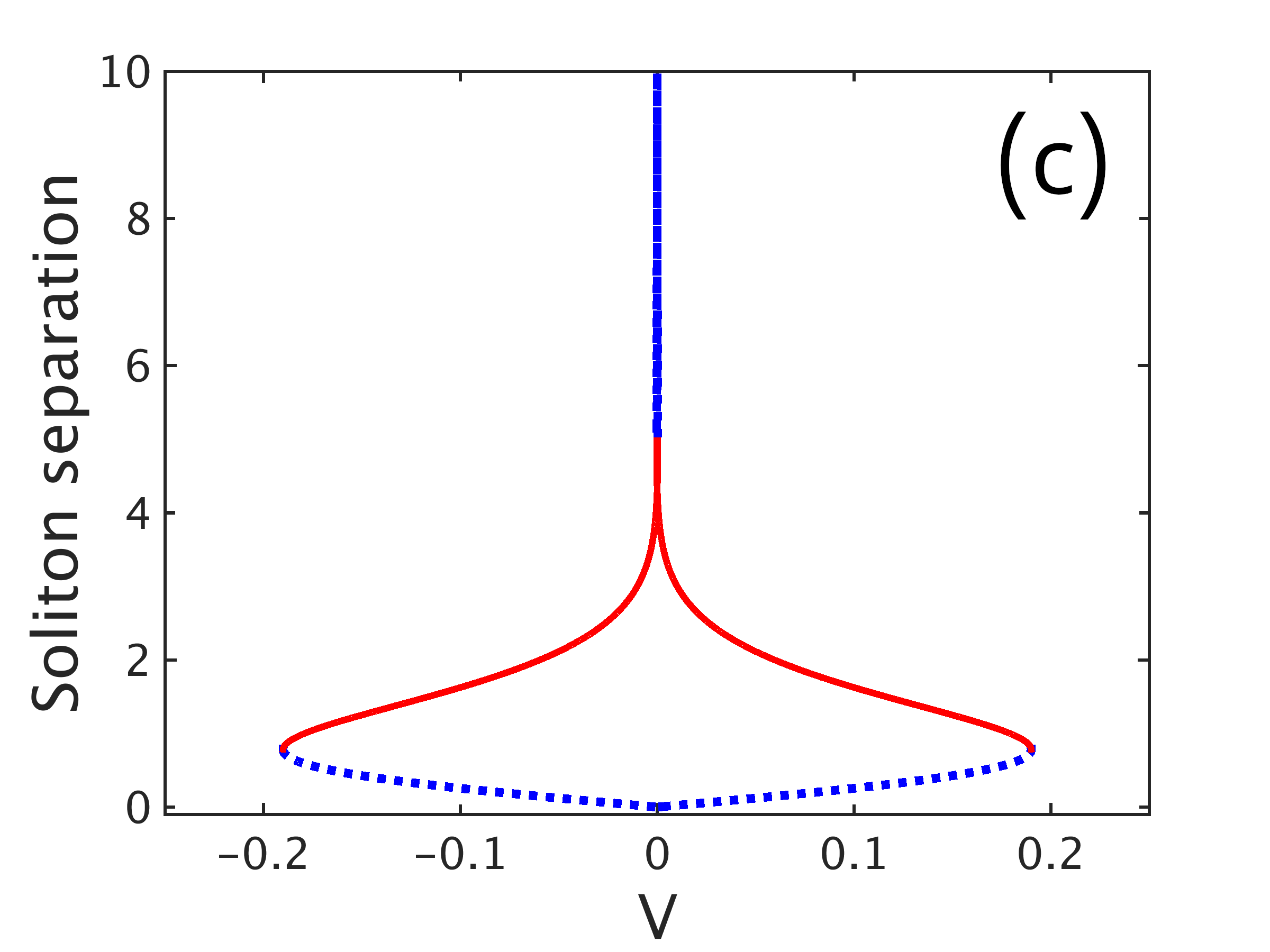}
  \end{minipage}
  \hfill
  \begin{minipage}{0.48\textwidth}
    \centering
    \includegraphics[width=\linewidth]{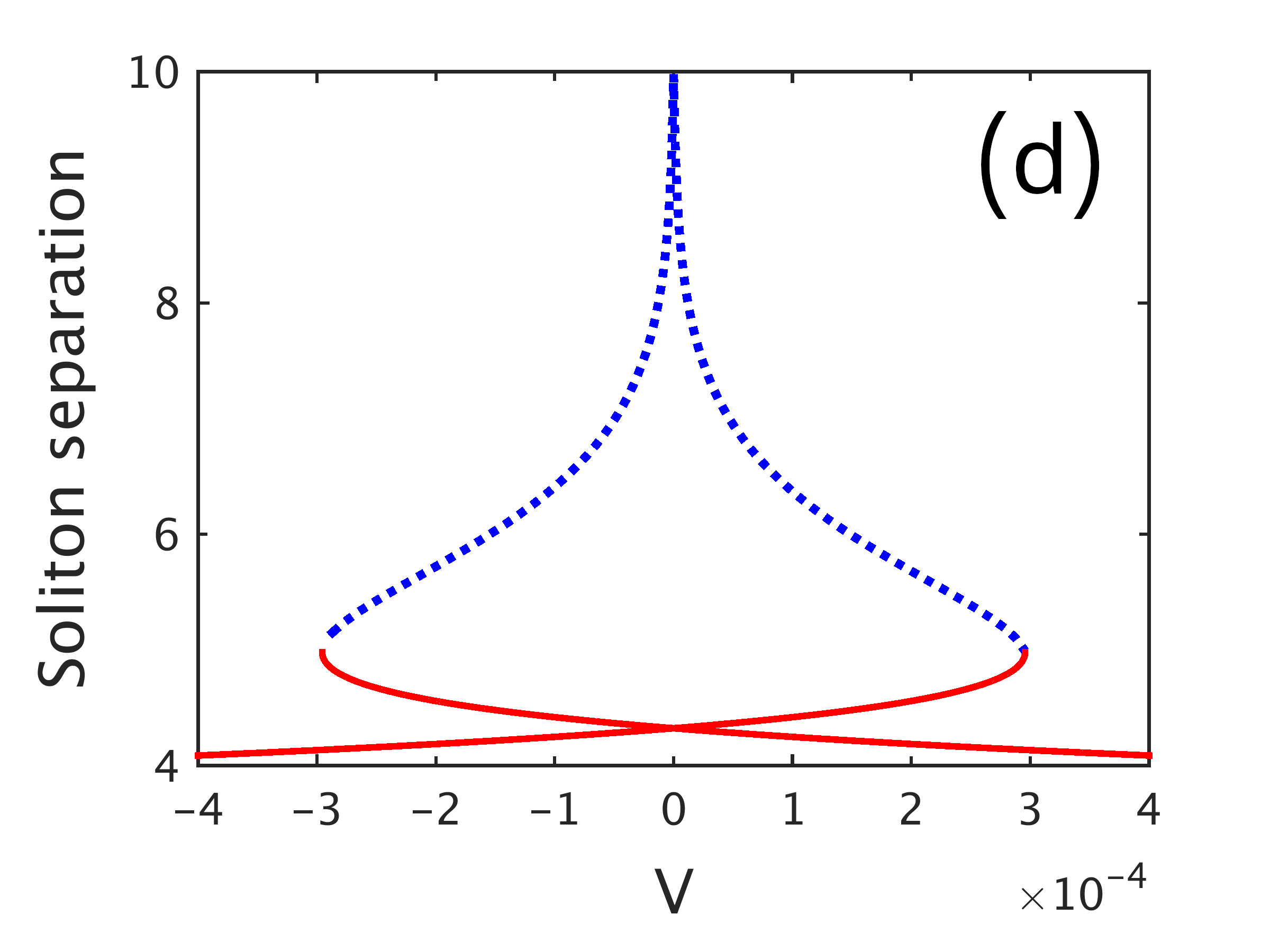}
  \end{minipage}
\caption{Soliton separation $|\xi_-|$ as a function of the cavity length mismatch parameter $V$. 
Stable and unstable solutions are shown by red solid and blue dotted lines, respectively.  
(a) The lower (upper) segment of the branch shown with red solid line corresponds to the stable bunching (anti-bunching) state, with $\delta = 0$.  
(b) The upper segment of the branch from (a) shown on a stretched $V$ scale.  
(c) The middle segment of the branch shown with red solid line corresponds to a stable soliton bound state, with $\delta = 2.64$.  
(d) The upper segment of the branch from (c) shown on a stretched $V$ scale.  
Other parameters: $\theta = 3.5$, $\eta_1 = \eta_2 = 1.9$, $b = 1$, and $L = 20.0$.}
  \label{fig:separation}
\end{figure}

Figure \ref{fig:separation} illustrates the dependence of the soliton separation on the cavity length mismatch parameter $V$. Panels (a) and (b) in this figure were calculated with zero injection phase  difference. The lower segment of the solution branch, shown by the red solid line in Fig.~\ref{fig:separation}(a), corresponds to the stable bunching state. It is seen that the soliton separation $|\xi_-|$ in this state increases with the absolute value of $V$ up to the saddle-node bifurcation points, where $\xi_-\approx 0.79$. This value approximately corresponds to half the distance between the maximum and minimum of the quantity $C_-$ shown in Fig.~\ref{fig:Cpm}. At these points, the stable bunching state merges with the unstable bound state solutions, which are shown by the blue dotted lines in Figs.~\ref{fig:separation}(a,b). 

\begin{figure}[htbp]
  \centering
  \begin{minipage}{0.48\textwidth}
    \centering
    \includegraphics[width=\linewidth]{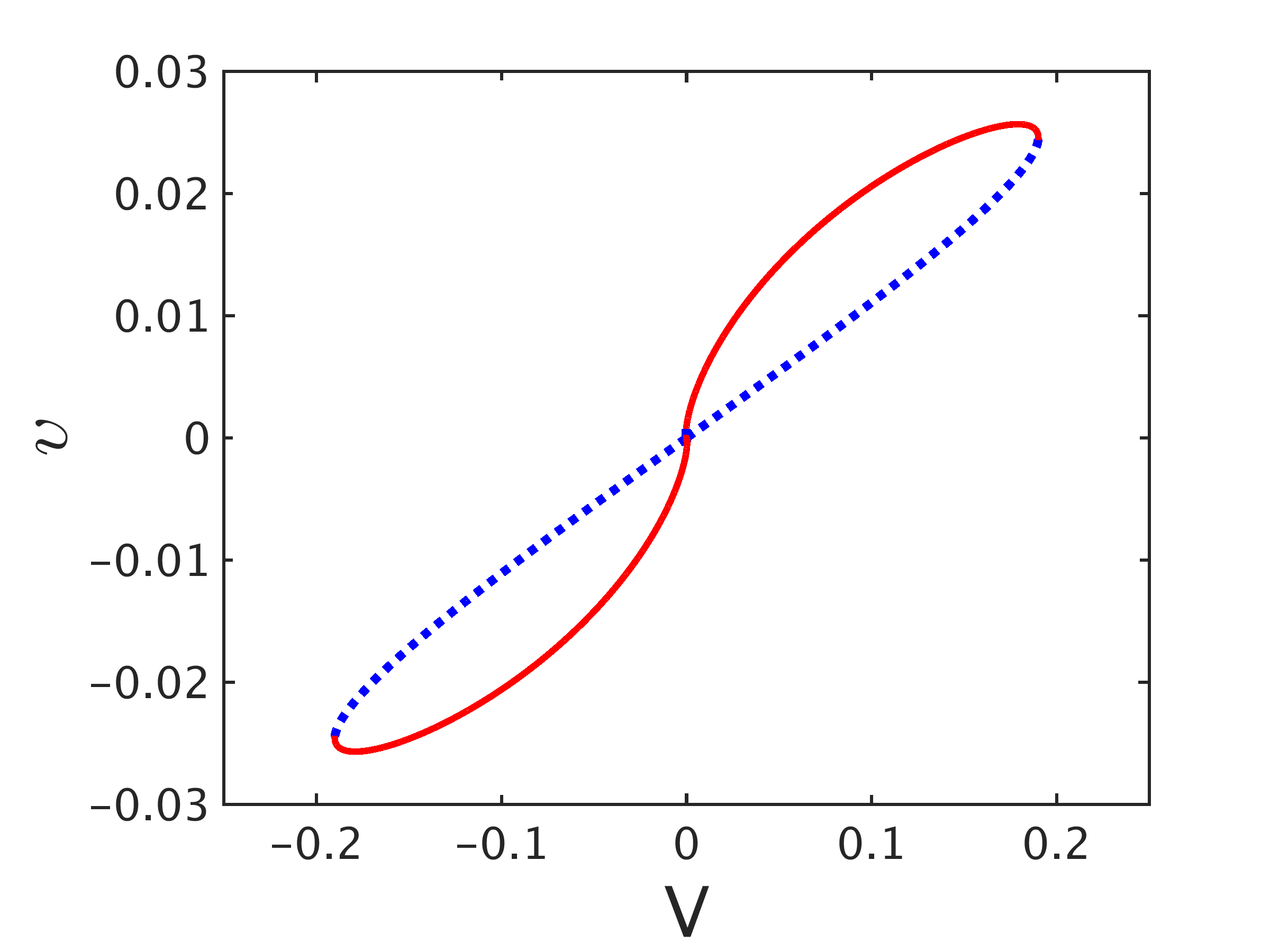}
  \end{minipage}
  \hfill
  \begin{minipage}{0.48\textwidth}
    \centering
    \includegraphics[width=\linewidth]{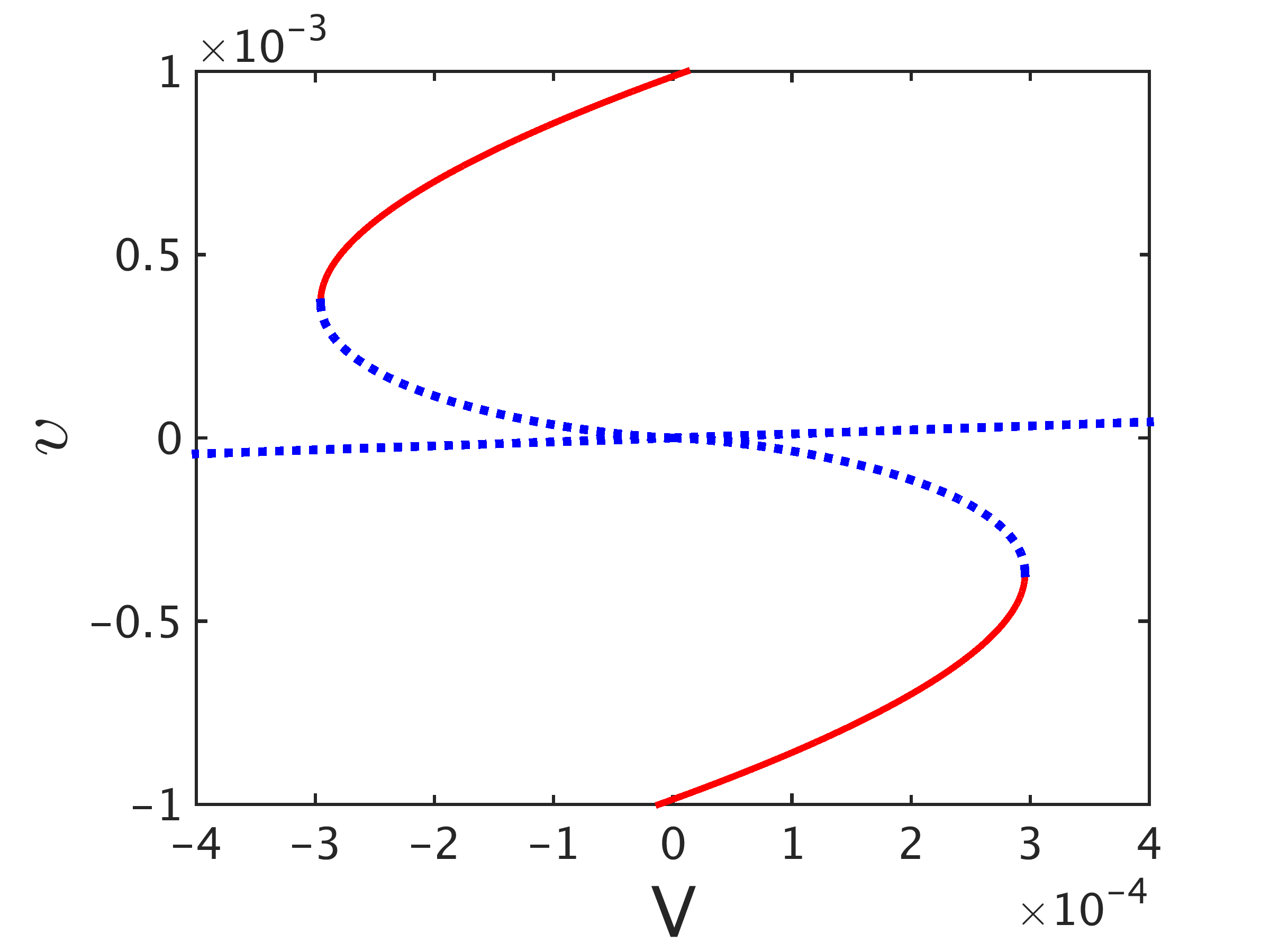}
  \end{minipage}
\caption{Soliton cluster drift velocity $\upsilon$ as function of the cavity length mismatch parameter $V$. 
Stable and unstable solutions are shown by red solid and blue dotted lines, respectively.  
(a) Two segments of the branch shown with red solid line correspond to the stable soliton bound state calculated at $\delta = 2.64$.  
(b) The upper segment of the branch from (a) shown on a stretched scale.   
Other parameters are as in Fig.~\ref{fig:separation}.}
  \label{fig:velocity}
\end{figure}

As shown in Fig.~\ref{fig:separation}(b), the two blue dotted lines intersect the $V=0$ axis at $|\xi_-| \approx 4.3$, which approximately corresponds to the soliton separation $\xi_- = \xi_-^*$ in the bound state at $V=0$, as predicted by our asymptotic analysis. With further increase of the soliton separation, the unstable bound state solutions merge with the stable anti-bunching states. These states are represented by the red solid line in Fig.~\ref{fig:separation}(b). It is evident that the cavity length mismatch interval where the anti-bunching state exists is much narrower than that of the bunching state. This results from the very weak interaction between solitons at their maximal possible separation.

Panels (c) and (d) in Fig.~\ref{fig:separation} are analogous to panels (a) and (b), but correspond to calculations with the injection phase difference $\delta = 2.64$. In this regime, bunching and anti-bunching soliton clusters are unstable, whereas soliton bound states are stable, as indicated by the red solid lines. Note that, unlike panels (a) and (b) — which are obtained for $\delta = 0$ and therefore, according to Eq.~(\ref{eq:int1}), correspond to zero drift velocity $\partial_t \xi_+ = v = 0$ — the solution branches in panels (c) and (d) generally exhibit a nonzero drift velocity. The dependence of this velocity on the cavity length mismatch parameter $V$ is illustrated in Fig.~\ref{fig:velocity}.

\section{Conclusion}
We analytically and numerically investigated soliton interactions in a system of two coupled Lugiato–Lefever equations (\ref{eq:LLE_coupled1})-(\ref{eq:LLE_coupled2}), which describe two nearly identical, weakly coupled cavities, each driven by its own injection beam. In the weak-coupling regime and in the absence of cavity length mismatch, asymptotic analysis of two-soliton interactions revealed three distinct types of soliton clusters, characterized by a constant difference in soliton peak positions, $\xi_-$. In addition to the bunching cluster with zero separation ($\xi_- = 0$), we identified a bound-state cluster originating from soliton tail interactions, in which the soliton separation depends on the soliton size but is almost independent of the cavity length. We also found an anti-bunching state with maximal separation $\xi_- = L/2$, where $L$ is the cavity length. The latter was determined to be only marginally stable and highly sensitive to asymmetric perturbations. Finally, depending on the relative phase of the pumping beams, either the bunching and anti-bunching clusters or the bound-state cluster can be stabilized.

The numerical simulations show excellent agreement with the asymptotic analysis. They reveal that, as predicted by asymptotic interaction equations, in the presence of a cavity length mismatch, solitons in the the bunching state acquire a finite soliton separation that grows with the mismatch, and finally the cluster is destroyed after the mismatch achieves the saddle node bifurcation threshold, where the bunching state merges with the bound state solution.  Likewise, changing the cavity length mismatch allows the soliton separation in both the bound-state and anti-bunching clusters to be tuned, but the anti-bunching cluster exists only over an extremely narrow range of mismatch. Remarkably, even for identical cavity lengths, a nonzero separation can emerge through a pitchfork bifurcation when the phase difference between the injection fields is appropriately tuned. These findings offer new insights into soliton control and highlight potential strategies for manipulating localized states in coupled cavity systems.
\clearpage

\begin{acknowledgments}
This work was supported by the DFG project No. 491234846, the RSF project No.~25-11-20069, the grant No. RPG-2021-072.
\end{acknowledgments}

 \bibliographystyle{apsrev4-2}
\bibliography{references}

\begin{thebibliography}{33}%
\makeatletter
\providecommand \@ifxundefined [1]{%
 \@ifx{#1\undefined}
}%
\providecommand \@ifnum [1]{%
 \ifnum #1\expandafter \@firstoftwo
 \else \expandafter \@secondoftwo
 \fi
}%
\providecommand \@ifx [1]{%
 \ifx #1\expandafter \@firstoftwo
 \else \expandafter \@secondoftwo
 \fi
}%
\providecommand \natexlab [1]{#1}%
\providecommand \enquote  [1]{``#1''}%
\providecommand \bibnamefont  [1]{#1}%
\providecommand \bibfnamefont [1]{#1}%
\providecommand \citenamefont [1]{#1}%
\providecommand \href@noop [0]{\@secondoftwo}%
\providecommand \href [0]{\begingroup \@sanitize@url \@href}%
\providecommand \@href[1]{\@@startlink{#1}\@@href}%
\providecommand \@@href[1]{\endgroup#1\@@endlink}%
\providecommand \@sanitize@url [0]{\catcode `\\12\catcode `\$12\catcode `\&12\catcode `\#12\catcode `\^12\catcode `\_12\catcode `\%12\relax}%
\providecommand \@@startlink[1]{}%
\providecommand \@@endlink[0]{}%
\providecommand \url  [0]{\begingroup\@sanitize@url \@url }%
\providecommand \@url [1]{\endgroup\@href {#1}{\urlprefix }}%
\providecommand \urlprefix  [0]{URL }%
\providecommand \Eprint [0]{\href }%
\providecommand \doibase [0]{https://doi.org/}%
\providecommand \selectlanguage [0]{\@gobble}%
\providecommand \bibinfo  [0]{\@secondoftwo}%
\providecommand \bibfield  [0]{\@secondoftwo}%
\providecommand \translation [1]{[#1]}%
\providecommand \BibitemOpen [0]{}%
\providecommand \bibitemStop [0]{}%
\providecommand \bibitemNoStop [0]{.\EOS\space}%
\providecommand \EOS [0]{\spacefactor3000\relax}%
\providecommand \BibitemShut  [1]{\csname bibitem#1\endcsname}%
\let\auto@bib@innerbib\@empty
\bibitem [{\citenamefont {Herr}\ \emph {et~al.}(2014{\natexlab{a}})\citenamefont {Herr}, \citenamefont {Brasch}, \citenamefont {Jost}, \citenamefont {Wang}, \citenamefont {Kondratiev}, \citenamefont {Gorodetsky},\ and\ \citenamefont {Kippenberg}}]{herr2014temporal}%
  \BibitemOpen
  \bibfield  {author} {\bibinfo {author} {\bibfnamefont {T.}~\bibnamefont {Herr}}, \bibinfo {author} {\bibfnamefont {V.}~\bibnamefont {Brasch}}, \bibinfo {author} {\bibfnamefont {J.~D.}\ \bibnamefont {Jost}}, \bibinfo {author} {\bibfnamefont {C.~Y.}\ \bibnamefont {Wang}}, \bibinfo {author} {\bibfnamefont {N.~M.}\ \bibnamefont {Kondratiev}}, \bibinfo {author} {\bibfnamefont {M.~L.}\ \bibnamefont {Gorodetsky}},\ and\ \bibinfo {author} {\bibfnamefont {T.~J.}\ \bibnamefont {Kippenberg}},\ }\href {https://doi.org/10.1038/nphoton.2013.343} {\bibfield  {journal} {\bibinfo  {journal} {Nature Photonics}\ }\textbf {\bibinfo {volume} {8}},\ \bibinfo {pages} {145} (\bibinfo {year} {2014}{\natexlab{a}})}\BibitemShut {NoStop}%
\bibitem [{\citenamefont {Kippenberg}\ \emph {et~al.}(2018)\citenamefont {Kippenberg}, \citenamefont {Gaeta}, \citenamefont {Lipson},\ and\ \citenamefont {Gorodetsky}}]{kippenberg2018dissipative}%
  \BibitemOpen
  \bibfield  {author} {\bibinfo {author} {\bibfnamefont {T.~J.}\ \bibnamefont {Kippenberg}}, \bibinfo {author} {\bibfnamefont {A.~L.}\ \bibnamefont {Gaeta}}, \bibinfo {author} {\bibfnamefont {M.}~\bibnamefont {Lipson}},\ and\ \bibinfo {author} {\bibfnamefont {M.~L.}\ \bibnamefont {Gorodetsky}},\ }\href {https://doi.org/10.1126/science.aan8083} {\bibfield  {journal} {\bibinfo  {journal} {Science}\ }\textbf {\bibinfo {volume} {361}},\ \bibinfo {pages} {eaan8083} (\bibinfo {year} {2018})}\BibitemShut {NoStop}%
\bibitem [{\citenamefont {Herr}\ \emph {et~al.}(2016)\citenamefont {Herr}, \citenamefont {Gorodetsky},\ and\ \citenamefont {Kippenberg}}]{herr2016dissipative}%
  \BibitemOpen
  \bibfield  {author} {\bibinfo {author} {\bibfnamefont {T.}~\bibnamefont {Herr}}, \bibinfo {author} {\bibfnamefont {M.~L.}\ \bibnamefont {Gorodetsky}},\ and\ \bibinfo {author} {\bibfnamefont {T.~J.}\ \bibnamefont {Kippenberg}},\ }in\ \href {https://doi.org/10.1002/9783527686476.ch6} {\emph {\bibinfo {booktitle} {Nonlinear Optical Cavity Dynamics: From Microresonators to Fiber Lasers}}},\ \bibinfo {editor} {edited by\ \bibinfo {editor} {\bibfnamefont {P.}~\bibnamefont {Grelu}}}\ (\bibinfo  {publisher} {Wiley‐VCH Verlag GmbH \& Co. KGaA},\ \bibinfo {year} {2016})\ Chap.~\bibinfo {chapter} {6}, pp.\ \bibinfo {pages} {157--179}\BibitemShut {NoStop}%
\bibitem [{\citenamefont {Xue}\ \emph {et~al.}(2019)\citenamefont {Xue}, \citenamefont {Zheng},\ and\ \citenamefont {Zhou}}]{xue2019super}%
  \BibitemOpen
  \bibfield  {author} {\bibinfo {author} {\bibfnamefont {X.}~\bibnamefont {Xue}}, \bibinfo {author} {\bibfnamefont {X.}~\bibnamefont {Zheng}},\ and\ \bibinfo {author} {\bibfnamefont {B.}~\bibnamefont {Zhou}},\ }\href@noop {} {\bibfield  {journal} {\bibinfo  {journal} {Nature Photonics}\ }\textbf {\bibinfo {volume} {13}},\ \bibinfo {pages} {616} (\bibinfo {year} {2019})}\BibitemShut {NoStop}%
\bibitem [{\citenamefont {Yuan}\ \emph {et~al.}(2023)\citenamefont {Yuan}, \citenamefont {Gao}, \citenamefont {Yu}, \citenamefont {Wang}, \citenamefont {Jin}, \citenamefont {Ji}, \citenamefont {Feshali}, \citenamefont {Paniccia}, \citenamefont {Bowers},\ and\ \citenamefont {Vahala}}]{yuan2023soliton}%
  \BibitemOpen
  \bibfield  {author} {\bibinfo {author} {\bibfnamefont {Z.}~\bibnamefont {Yuan}}, \bibinfo {author} {\bibfnamefont {M.}~\bibnamefont {Gao}}, \bibinfo {author} {\bibfnamefont {Y.}~\bibnamefont {Yu}}, \bibinfo {author} {\bibfnamefont {H.}~\bibnamefont {Wang}}, \bibinfo {author} {\bibfnamefont {W.}~\bibnamefont {Jin}}, \bibinfo {author} {\bibfnamefont {Q.-X.}\ \bibnamefont {Ji}}, \bibinfo {author} {\bibfnamefont {A.}~\bibnamefont {Feshali}}, \bibinfo {author} {\bibfnamefont {M.}~\bibnamefont {Paniccia}}, \bibinfo {author} {\bibfnamefont {J.}~\bibnamefont {Bowers}},\ and\ \bibinfo {author} {\bibfnamefont {K.}~\bibnamefont {Vahala}},\ }\href@noop {} {\bibfield  {journal} {\bibinfo  {journal} {Nature Photonics}\ }\textbf {\bibinfo {volume} {17}},\ \bibinfo {pages} {977} (\bibinfo {year} {2023})}\BibitemShut {NoStop}%
\bibitem [{\citenamefont {Aranson}\ and\ \citenamefont {Kramer}(2002)}]{Aranson2002}%
  \BibitemOpen
  \bibfield  {author} {\bibinfo {author} {\bibfnamefont {I.~S.}\ \bibnamefont {Aranson}}\ and\ \bibinfo {author} {\bibfnamefont {L.}~\bibnamefont {Kramer}},\ }\href {https://doi.org/10.1103/RevModPhys.74.99} {\bibfield  {journal} {\bibinfo  {journal} {Reviews of Modern Physics}\ }\textbf {\bibinfo {volume} {74}},\ \bibinfo {pages} {99} (\bibinfo {year} {2002})}\BibitemShut {NoStop}%
\bibitem [{\citenamefont {Lugiato}\ and\ \citenamefont {Lefever}(1987{\natexlab{a}})}]{Lugiato1987}%
  \BibitemOpen
  \bibfield  {author} {\bibinfo {author} {\bibfnamefont {L.~A.}\ \bibnamefont {Lugiato}}\ and\ \bibinfo {author} {\bibfnamefont {R.}~\bibnamefont {Lefever}},\ }\href {https://doi.org/10.1103/PhysRevLett.58.2209} {\bibfield  {journal} {\bibinfo  {journal} {Physical Review Letters}\ }\textbf {\bibinfo {volume} {58}},\ \bibinfo {pages} {2209} (\bibinfo {year} {1987}{\natexlab{a}})}\BibitemShut {NoStop}%
\bibitem [{\citenamefont {Akhmediev}\ \emph {et~al.}(1997)\citenamefont {Akhmediev}, \citenamefont {Ankiewicz},\ and\ \citenamefont {Soto-Crespo}}]{Akhmediev1997}%
  \BibitemOpen
  \bibfield  {author} {\bibinfo {author} {\bibfnamefont {N.}~\bibnamefont {Akhmediev}}, \bibinfo {author} {\bibfnamefont {A.}~\bibnamefont {Ankiewicz}},\ and\ \bibinfo {author} {\bibfnamefont {J.}~\bibnamefont {Soto-Crespo}},\ }\href@noop {} {\bibfield  {journal} {\bibinfo  {journal} {Physical Review Letters}\ }\textbf {\bibinfo {volume} {79}},\ \bibinfo {pages} {4047} (\bibinfo {year} {1997})}\BibitemShut {NoStop}%
\bibitem [{\citenamefont {Akhmediev}\ \emph {et~al.}(2005)\citenamefont {Akhmediev}, \citenamefont {Soto-Crespo}, \citenamefont {Grapinet},\ and\ \citenamefont {Grelu}}]{akhmediev2005dissipative}%
  \BibitemOpen
  \bibfield  {author} {\bibinfo {author} {\bibfnamefont {N.}~\bibnamefont {Akhmediev}}, \bibinfo {author} {\bibfnamefont {J.~M.}\ \bibnamefont {Soto-Crespo}}, \bibinfo {author} {\bibfnamefont {M.}~\bibnamefont {Grapinet}},\ and\ \bibinfo {author} {\bibfnamefont {P.}~\bibnamefont {Grelu}},\ }\href@noop {} {\bibfield  {journal} {\bibinfo  {journal} {Optical Fiber Technology}\ }\textbf {\bibinfo {volume} {11}},\ \bibinfo {pages} {209} (\bibinfo {year} {2005})}\BibitemShut {NoStop}%
\bibitem [{\citenamefont {Soto-Crespo}\ \emph {et~al.}(2007)\citenamefont {Soto-Crespo}, \citenamefont {Grelu}, \citenamefont {Akhmediev},\ and\ \citenamefont {Devine}}]{soto2007soliton}%
  \BibitemOpen
  \bibfield  {author} {\bibinfo {author} {\bibfnamefont {J.~M.}\ \bibnamefont {Soto-Crespo}}, \bibinfo {author} {\bibfnamefont {P.}~\bibnamefont {Grelu}}, \bibinfo {author} {\bibfnamefont {N.}~\bibnamefont {Akhmediev}},\ and\ \bibinfo {author} {\bibfnamefont {N.}~\bibnamefont {Devine}},\ }\href@noop {} {\bibfield  {journal} {\bibinfo  {journal} {Physical Review E—Statistical, Nonlinear, and Soft Matter Physics}\ }\textbf {\bibinfo {volume} {75}},\ \bibinfo {pages} {016613} (\bibinfo {year} {2007})}\BibitemShut {NoStop}%
\bibitem [{\citenamefont {Grelu}\ and\ \citenamefont {Akhmediev}(2012)}]{Grelu2012}%
  \BibitemOpen
  \bibfield  {author} {\bibinfo {author} {\bibfnamefont {P.}~\bibnamefont {Grelu}}\ and\ \bibinfo {author} {\bibfnamefont {N.}~\bibnamefont {Akhmediev}},\ }\href {https://doi.org/10.1038/nphoton.2011.301} {\bibfield  {journal} {\bibinfo  {journal} {Nature Photonics}\ }\textbf {\bibinfo {volume} {6}},\ \bibinfo {pages} {84} (\bibinfo {year} {2012})}\BibitemShut {NoStop}%
\bibitem [{\citenamefont {Vladimirov}\ \emph {et~al.}(1999)\citenamefont {Vladimirov}, \citenamefont {Fedorov}, \citenamefont {Kaliteevskii}, \citenamefont {Khodova},\ and\ \citenamefont {Rosanov}}]{vladimirov1999numerical}%
  \BibitemOpen
  \bibfield  {author} {\bibinfo {author} {\bibfnamefont {A.}~\bibnamefont {Vladimirov}}, \bibinfo {author} {\bibfnamefont {S.}~\bibnamefont {Fedorov}}, \bibinfo {author} {\bibfnamefont {N.}~\bibnamefont {Kaliteevskii}}, \bibinfo {author} {\bibfnamefont {G.}~\bibnamefont {Khodova}},\ and\ \bibinfo {author} {\bibfnamefont {N.}~\bibnamefont {Rosanov}},\ }\href@noop {} {\bibfield  {journal} {\bibinfo  {journal} {Journal of Optics B: Quantum and Semiclassical Optics}\ }\textbf {\bibinfo {volume} {1}},\ \bibinfo {pages} {101} (\bibinfo {year} {1999})}\BibitemShut {NoStop}%
\bibitem [{\citenamefont {Parra-Rivas}\ \emph {et~al.}(2014)\citenamefont {Parra-Rivas}, \citenamefont {Gomila}, \citenamefont {Mat{\'\i}as}, \citenamefont {Coen},\ and\ \citenamefont {Gelens}}]{parra2014dynamics}%
  \BibitemOpen
  \bibfield  {author} {\bibinfo {author} {\bibfnamefont {P.}~\bibnamefont {Parra-Rivas}}, \bibinfo {author} {\bibfnamefont {D.}~\bibnamefont {Gomila}}, \bibinfo {author} {\bibfnamefont {M.~A.}\ \bibnamefont {Mat{\'\i}as}}, \bibinfo {author} {\bibfnamefont {S.}~\bibnamefont {Coen}},\ and\ \bibinfo {author} {\bibfnamefont {L.}~\bibnamefont {Gelens}},\ }\href@noop {} {\bibfield  {journal} {\bibinfo  {journal} {Physical Review A}\ }\textbf {\bibinfo {volume} {89}},\ \bibinfo {pages} {043813} (\bibinfo {year} {2014})}\BibitemShut {NoStop}%
\bibitem [{\citenamefont {Parra-Rivas}\ \emph {et~al.}(2017)\citenamefont {Parra-Rivas}, \citenamefont {Gomila}, \citenamefont {Colet},\ and\ \citenamefont {Gelens}}]{parra2017interaction}%
  \BibitemOpen
  \bibfield  {author} {\bibinfo {author} {\bibfnamefont {P.}~\bibnamefont {Parra-Rivas}}, \bibinfo {author} {\bibfnamefont {D.}~\bibnamefont {Gomila}}, \bibinfo {author} {\bibfnamefont {P.}~\bibnamefont {Colet}},\ and\ \bibinfo {author} {\bibfnamefont {L.}~\bibnamefont {Gelens}},\ }\href@noop {} {\bibfield  {journal} {\bibinfo  {journal} {The European Physical Journal D}\ }\textbf {\bibinfo {volume} {71}},\ \bibinfo {pages} {198} (\bibinfo {year} {2017})}\BibitemShut {NoStop}%
\bibitem [{\citenamefont {Vladimirov}\ \emph {et~al.}(2021)\citenamefont {Vladimirov}, \citenamefont {Tlidi},\ and\ \citenamefont {Taki}}]{vladimirov2021dissipative}%
  \BibitemOpen
  \bibfield  {author} {\bibinfo {author} {\bibfnamefont {A.~G.}\ \bibnamefont {Vladimirov}}, \bibinfo {author} {\bibfnamefont {M.}~\bibnamefont {Tlidi}},\ and\ \bibinfo {author} {\bibfnamefont {M.}~\bibnamefont {Taki}},\ }\href@noop {} {\bibfield  {journal} {\bibinfo  {journal} {Physical Review A}\ }\textbf {\bibinfo {volume} {103}},\ \bibinfo {pages} {063505} (\bibinfo {year} {2021})}\BibitemShut {NoStop}%
\bibitem [{\citenamefont {Nizette}\ \emph {et~al.}(2006)\citenamefont {Nizette}, \citenamefont {Rachinskii}, \citenamefont {Vladimirov},\ and\ \citenamefont {Wolfrum}}]{nizette2006pulse}%
  \BibitemOpen
  \bibfield  {author} {\bibinfo {author} {\bibfnamefont {M.}~\bibnamefont {Nizette}}, \bibinfo {author} {\bibfnamefont {D.}~\bibnamefont {Rachinskii}}, \bibinfo {author} {\bibfnamefont {A.}~\bibnamefont {Vladimirov}},\ and\ \bibinfo {author} {\bibfnamefont {M.}~\bibnamefont {Wolfrum}},\ }\href@noop {} {\bibfield  {journal} {\bibinfo  {journal} {Physica D: Nonlinear Phenomena}\ }\textbf {\bibinfo {volume} {218}},\ \bibinfo {pages} {95} (\bibinfo {year} {2006})}\BibitemShut {NoStop}%
\bibitem [{\citenamefont {Javaloyes}\ \emph {et~al.}(2016)\citenamefont {Javaloyes}, \citenamefont {Camelin}, \citenamefont {Marconi},\ and\ \citenamefont {Giudici}}]{javaloyes2016dynamics}%
  \BibitemOpen
  \bibfield  {author} {\bibinfo {author} {\bibfnamefont {J.}~\bibnamefont {Javaloyes}}, \bibinfo {author} {\bibfnamefont {P.}~\bibnamefont {Camelin}}, \bibinfo {author} {\bibfnamefont {M.}~\bibnamefont {Marconi}},\ and\ \bibinfo {author} {\bibfnamefont {M.}~\bibnamefont {Giudici}},\ }\href@noop {} {\bibfield  {journal} {\bibinfo  {journal} {Physical Review Letters}\ }\textbf {\bibinfo {volume} {116}},\ \bibinfo {pages} {133901} (\bibinfo {year} {2016})}\BibitemShut {NoStop}%
\bibitem [{\citenamefont {Vladimirov}(2022)}]{vladimirov2022short}%
  \BibitemOpen
  \bibfield  {author} {\bibinfo {author} {\bibfnamefont {A.~G.}\ \bibnamefont {Vladimirov}},\ }\href@noop {} {\bibfield  {journal} {\bibinfo  {journal} {Physical Review E}\ }\textbf {\bibinfo {volume} {105}},\ \bibinfo {pages} {044207} (\bibinfo {year} {2022})}\BibitemShut {NoStop}%
\bibitem [{\citenamefont {Vladimirov}(2023)}]{vladimirov2023temporal}%
  \BibitemOpen
  \bibfield  {author} {\bibinfo {author} {\bibfnamefont {A.~G.}\ \bibnamefont {Vladimirov}},\ }\href@noop {} {\bibfield  {journal} {\bibinfo  {journal} {Optics}\ }\textbf {\bibinfo {volume} {4}},\ \bibinfo {pages} {433} (\bibinfo {year} {2023})}\BibitemShut {NoStop}%
\bibitem [{\citenamefont {Malomed}(1991)}]{Malomed1991}%
  \BibitemOpen
  \bibfield  {author} {\bibinfo {author} {\bibfnamefont {B.~A.}\ \bibnamefont {Malomed}},\ }\href {https://doi.org/10.1103/PhysRevA.44.6954} {\bibfield  {journal} {\bibinfo  {journal} {Physical Review A}\ }\textbf {\bibinfo {volume} {44}},\ \bibinfo {pages} {6954} (\bibinfo {year} {1991})}\BibitemShut {NoStop}%
\bibitem [{\citenamefont {Vladimirov}\ \emph {et~al.}(2001)\citenamefont {Vladimirov}, \citenamefont {Khodova},\ and\ \citenamefont {Rosanov}}]{vladimirov2001stable}%
  \BibitemOpen
  \bibfield  {author} {\bibinfo {author} {\bibfnamefont {A.}~\bibnamefont {Vladimirov}}, \bibinfo {author} {\bibfnamefont {G.}~\bibnamefont {Khodova}},\ and\ \bibinfo {author} {\bibfnamefont {N.}~\bibnamefont {Rosanov}},\ }\href@noop {} {\bibfield  {journal} {\bibinfo  {journal} {Physical Review E}\ }\textbf {\bibinfo {volume} {63}},\ \bibinfo {pages} {056607} (\bibinfo {year} {2001})}\BibitemShut {NoStop}%
\bibitem [{\citenamefont {Descalzi}\ \emph {et~al.}(2011)\citenamefont {Descalzi}, \citenamefont {Clerc}, \citenamefont {Residori},\ and\ \citenamefont {Assanto}}]{Descalzi2011}%
  \BibitemOpen
  \bibinfo {editor} {\bibfnamefont {O.}~\bibnamefont {Descalzi}}, \bibinfo {editor} {\bibfnamefont {M.}~\bibnamefont {Clerc}}, \bibinfo {editor} {\bibfnamefont {S.}~\bibnamefont {Residori}},\ and\ \bibinfo {editor} {\bibfnamefont {G.}~\bibnamefont {Assanto}},\ eds.,\ \href {https://doi.org/10.1007/978-3-642-16549-2} {\emph {\bibinfo {title} {Localized States in Physics: Solitons and Patterns}}},\ \bibinfo {edition} {1st}\ ed.\ (\bibinfo  {publisher} {Springer},\ \bibinfo {address} {Berlin, Heidelberg},\ \bibinfo {year} {2011})\ p.\ \bibinfo {pages} {286}\BibitemShut {NoStop}%
\bibitem [{\citenamefont {Turaev}\ \emph {et~al.}(2012)\citenamefont {Turaev}, \citenamefont {Vladimirov},\ and\ \citenamefont {Zelik}}]{turaev2012long}%
  \BibitemOpen
  \bibfield  {author} {\bibinfo {author} {\bibfnamefont {D.}~\bibnamefont {Turaev}}, \bibinfo {author} {\bibfnamefont {A.}~\bibnamefont {Vladimirov}},\ and\ \bibinfo {author} {\bibfnamefont {S.}~\bibnamefont {Zelik}},\ }\href@noop {} {\bibfield  {journal} {\bibinfo  {journal} {Physical Review Letters}\ }\textbf {\bibinfo {volume} {108}},\ \bibinfo {pages} {263906} (\bibinfo {year} {2012})}\BibitemShut {NoStop}%
\bibitem [{\citenamefont {Tang}\ \emph {et~al.}(2005)\citenamefont {Tang}, \citenamefont {Zhang}, \citenamefont {Zhao},\ and\ \citenamefont {Wu}}]{Tang2005}%
  \BibitemOpen
  \bibfield  {author} {\bibinfo {author} {\bibfnamefont {D.~Y.}\ \bibnamefont {Tang}}, \bibinfo {author} {\bibfnamefont {H.}~\bibnamefont {Zhang}}, \bibinfo {author} {\bibfnamefont {L.~M.}\ \bibnamefont {Zhao}},\ and\ \bibinfo {author} {\bibfnamefont {X.}~\bibnamefont {Wu}},\ }\href {https://doi.org/10.1103/PhysRevLett.101.153904} {\bibfield  {journal} {\bibinfo  {journal} {Physical Review Letters}\ }\textbf {\bibinfo {volume} {101}},\ \bibinfo {pages} {153904} (\bibinfo {year} {2005})}\BibitemShut {NoStop}%
\bibitem [{\citenamefont {Runge}\ \emph {et~al.}(2015)\citenamefont {Runge}, \citenamefont {Broderick},\ and\ \citenamefont {Erkintalo}}]{Runge2015}%
  \BibitemOpen
  \bibfield  {author} {\bibinfo {author} {\bibfnamefont {A.~F.~J.}\ \bibnamefont {Runge}}, \bibinfo {author} {\bibfnamefont {N.~G.~R.}\ \bibnamefont {Broderick}},\ and\ \bibinfo {author} {\bibfnamefont {M.}~\bibnamefont {Erkintalo}},\ }\href {https://doi.org/10.1364/OPTICA.2.000036} {\bibfield  {journal} {\bibinfo  {journal} {Optica}\ }\textbf {\bibinfo {volume} {2}},\ \bibinfo {pages} {36} (\bibinfo {year} {2015})}\BibitemShut {NoStop}%
\bibitem [{\citenamefont {Herr}\ \emph {et~al.}(2014{\natexlab{b}})\citenamefont {Herr}, \citenamefont {Brasch}, \citenamefont {Jost}, \citenamefont {F\"ul\"op}, \citenamefont {Kondratiev}, \citenamefont {Gorodetsky},\ and\ \citenamefont {Kippenberg}}]{Herr2014}%
  \BibitemOpen
  \bibfield  {author} {\bibinfo {author} {\bibfnamefont {T.}~\bibnamefont {Herr}}, \bibinfo {author} {\bibfnamefont {V.}~\bibnamefont {Brasch}}, \bibinfo {author} {\bibfnamefont {J.~D.}\ \bibnamefont {Jost}}, \bibinfo {author} {\bibfnamefont {C.~W. H.~P.}\ \bibnamefont {F\"ul\"op}}, \bibinfo {author} {\bibfnamefont {N.~M.}\ \bibnamefont {Kondratiev}}, \bibinfo {author} {\bibfnamefont {M.~L.}\ \bibnamefont {Gorodetsky}},\ and\ \bibinfo {author} {\bibfnamefont {T.~J.}\ \bibnamefont {Kippenberg}},\ }\href {https://doi.org/10.1038/nphoton.2013.343} {\bibfield  {journal} {\bibinfo  {journal} {Nature Photonics}\ }\textbf {\bibinfo {volume} {8}},\ \bibinfo {pages} {145} (\bibinfo {year} {2014}{\natexlab{b}})}\BibitemShut {NoStop}%
\bibitem [{\citenamefont {Vladimirov}\ \emph {et~al.}(2018)\citenamefont {Vladimirov}, \citenamefont {Gurevich},\ and\ \citenamefont {Tlidi}}]{vladimirov2018effect}%
  \BibitemOpen
  \bibfield  {author} {\bibinfo {author} {\bibfnamefont {A.~G.}\ \bibnamefont {Vladimirov}}, \bibinfo {author} {\bibfnamefont {S.~V.}\ \bibnamefont {Gurevich}},\ and\ \bibinfo {author} {\bibfnamefont {M.}~\bibnamefont {Tlidi}},\ }\href@noop {} {\bibfield  {journal} {\bibinfo  {journal} {Physical Review A}\ }\textbf {\bibinfo {volume} {97}},\ \bibinfo {pages} {013816} (\bibinfo {year} {2018})}\BibitemShut {NoStop}%
\bibitem [{\citenamefont {Lugiato}\ and\ \citenamefont {Lefever}(1987{\natexlab{b}})}]{lugiato1987spatial}%
  \BibitemOpen
  \bibfield  {author} {\bibinfo {author} {\bibfnamefont {L.~A.}\ \bibnamefont {Lugiato}}\ and\ \bibinfo {author} {\bibfnamefont {R.}~\bibnamefont {Lefever}},\ }\href {https://doi.org/10.1103/PhysRevLett.58.2209} {\bibfield  {journal} {\bibinfo  {journal} {Physical Review Letters}\ }\textbf {\bibinfo {volume} {58}},\ \bibinfo {pages} {2209} (\bibinfo {year} {1987}{\natexlab{b}})}\BibitemShut {NoStop}%
\bibitem [{\citenamefont {Chembo}\ and\ \citenamefont {Menyuk}(2013)}]{chembo2013spatiotemporal}%
  \BibitemOpen
  \bibfield  {author} {\bibinfo {author} {\bibfnamefont {Y.~K.}\ \bibnamefont {Chembo}}\ and\ \bibinfo {author} {\bibfnamefont {C.~R.}\ \bibnamefont {Menyuk}},\ }\href {https://doi.org/10.1103/PhysRevA.87.053852} {\bibfield  {journal} {\bibinfo  {journal} {Physical Review A}\ }\textbf {\bibinfo {volume} {87}},\ \bibinfo {pages} {053852} (\bibinfo {year} {2013})}\BibitemShut {NoStop}%
\bibitem [{\citenamefont {Vladimirov}\ and\ \citenamefont {Dolinina}(2024)}]{vladimirov2023neutral}%
  \BibitemOpen
  \bibfield  {author} {\bibinfo {author} {\bibfnamefont {A.~G.}\ \bibnamefont {Vladimirov}}\ and\ \bibinfo {author} {\bibfnamefont {D.~A.}\ \bibnamefont {Dolinina}},\ }\href {https://doi.org/10.1103/PhysRevE.109.024206} {\bibfield  {journal} {\bibinfo  {journal} {Physical Review E}\ }\textbf {\bibinfo {volume} {109}},\ \bibinfo {pages} {024206} (\bibinfo {year} {2024})}\BibitemShut {NoStop}%
\bibitem [{\citenamefont {Tikan}\ \emph {et~al.}(2021)\citenamefont {Tikan}, \citenamefont {Anderson}, \citenamefont {Coen}, \citenamefont {Erkintalo}, \citenamefont {Gorodetsky}, \citenamefont {Kippenberg},\ and\ \citenamefont {Lucas}}]{tikan2021emergent}%
  \BibitemOpen
  \bibfield  {author} {\bibinfo {author} {\bibfnamefont {A.}~\bibnamefont {Tikan}}, \bibinfo {author} {\bibfnamefont {M.}~\bibnamefont {Anderson}}, \bibinfo {author} {\bibfnamefont {S.}~\bibnamefont {Coen}}, \bibinfo {author} {\bibfnamefont {M.}~\bibnamefont {Erkintalo}}, \bibinfo {author} {\bibfnamefont {M.~L.}\ \bibnamefont {Gorodetsky}}, \bibinfo {author} {\bibfnamefont {T.~J.}\ \bibnamefont {Kippenberg}},\ and\ \bibinfo {author} {\bibfnamefont {E.}~\bibnamefont {Lucas}},\ }\href {https://doi.org/10.1038/s41567-020-01138-x} {\bibfield  {journal} {\bibinfo  {journal} {Nature Physics}\ }\textbf {\bibinfo {volume} {17}},\ \bibinfo {pages} {604} (\bibinfo {year} {2021})}\BibitemShut {NoStop}%
\bibitem [{\citenamefont {Arnol'd}(1963)}]{arnol1963small}%
  \BibitemOpen
  \bibfield  {author} {\bibinfo {author} {\bibfnamefont {V.~I.}\ \bibnamefont {Arnol'd}},\ }\href@noop {} {\bibfield  {journal} {\bibinfo  {journal} {Russian Mathematical Surveys}\ }\textbf {\bibinfo {volume} {18}},\ \bibinfo {pages} {85} (\bibinfo {year} {1963})}\BibitemShut {NoStop}%
\bibitem [{\citenamefont {Pikovsky}\ \emph {et~al.}(2001)\citenamefont {Pikovsky}, \citenamefont {Rosenblum},\ and\ \citenamefont {Kurths}}]{pikovsky2001universal}%
  \BibitemOpen
  \bibfield  {author} {\bibinfo {author} {\bibfnamefont {A.}~\bibnamefont {Pikovsky}}, \bibinfo {author} {\bibfnamefont {M.}~\bibnamefont {Rosenblum}},\ and\ \bibinfo {author} {\bibfnamefont {J.}~\bibnamefont {Kurths}},\ }\href@noop {} {\emph {\bibinfo {title} {Synchronization. A universal concept in nonlinear sciences}}},\ \bibinfo {series} {Cambridge Nonlinear Science Series}, Vol.~\bibinfo {volume} {12}\ (\bibinfo  {publisher} {Cambridge University Press},\ \bibinfo {address} {New York},\ \bibinfo {year} {2001})\ pp.\ \bibinfo {pages} {10--1017}\BibitemShut {NoStop}%
\end{thebibliography}%
\end{document}